\documentclass[acmsmall,screen,nonacm]{acmart}
\usepackage{amsmath}
\usepackage{enumitem}
\usepackage{algorithmic}
\usepackage{graphicx}
\usepackage{textcomp}
\usepackage{xcolor}
\usepackage{colortbl} 
\definecolor{bestgreen}{RGB}{224,241,214}

\newcommand{\bestcell}[1]{%
  \cellcolor{bestgreen}#1%
}

\usepackage{booktabs}
\usepackage{threeparttable}
\usepackage{multirow}
\usepackage{siunitx}
\usepackage[most]{tcolorbox}
\usepackage{makecell}
\usepackage{xstring}
\usepackage{listings}
\usepackage{flafter}

\newtcolorbox{findingbox}{colback=gray!8, colframe=gray!50,
  boxrule=0.5pt, arc=2pt, left=4pt, right=4pt, top=2pt, bottom=2pt,
  before skip=5pt, after skip=5pt}

\definecolor{mockkeyword}{HTML}{C00040}
\definecolor{mockcomment}{HTML}{008C99}
\definecolor{mockstring}{HTML}{C87824}
\definecolor{mockhighlight}{HTML}{FCF3D2}
\newcommand{\mocklinebackground}{%
  \rlap{\kern8pt{\color{mockhighlight}%
    \rule[-\dp\strutbox]{\linewidth}{\dimexpr\ht\strutbox+\dp\strutbox\relax}}}}
\newcommand{\mocklinenumber}[1]{%
  {\tiny\color{gray}#1}%
  \ifnum\value{lstnumber}=13\mocklinebackground
  \else\ifnum\value{lstnumber}>3
    \ifnum\value{lstnumber}<8\mocklinebackground\fi
  \fi\fi}
\lstdefinestyle{mockexample}{
  language=Java,
  basicstyle=\scriptsize\ttfamily,
  keywordstyle=\color{mockkeyword}\bfseries,
  commentstyle=\color{mockcomment},
  stringstyle=\color{mockstring},
  numbers=left,
  numberstyle=\mocklinenumber,
  numbersep=8pt,
  xleftmargin=2.5em,
  frame=tb,
  framerule=0.4pt,
  rulecolor=\color{black!65},
  framesep=4pt,
  captionpos=b,
  showstringspaces=false,
  keepspaces=true,
  columns=fullflexible,
  breaklines=true,
  postbreak=\mbox{\textcolor{gray}{$\hookrightarrow$}\space}
}

\newcommand{\sd}[1]{{\tiny$\pm$#1}}

\def\BibTeX{{\rm B\kern-.05em{\sc i\kern-.025em b}\kern-.08em
    T\kern-.1667em\lower.7ex\hbox{E}\kern-.125emX}}

\usepackage{xcolor}
\usepackage{listings}
\usepackage{caption}

\definecolor{bugcolor}{HTML}{A83232}
\definecolor{goodcolor}{HTML}{247148}

\lstdefinestyle{mockcompare}{
  style=mockexample,
  frame=none,
  basicstyle=\fontsize{7}{9}\selectfont\ttfamily,
  showstringspaces=false,
  keepspaces=true,
  columns=fullflexible,
  breaklines=true,
  aboveskip=3pt,
  belowskip=3pt
}

\lstdefinestyle{mockcomparemethod}{
  style=mockcompare,
  basicstyle=\fontsize{6}{8}\selectfont\ttfamily,
  numberstyle=\tiny\color{gray},
  numbersep=5pt,
  xleftmargin=2em
}

\begin{document}
\raggedbottom

\title{Executable but Unfaithful: Mock Object, Stub, and Assertion Generation by LLMs and Coding Agents}

\author{Hao Ning}
\affiliation{\institution{Shandong University}\city{Qingdao}\country{China}}
\email{hning@mail.sdu.edu.cn}

\author{Jiahui Dong}
\affiliation{\institution{Shandong University}\city{Qingdao}\country{China}}
\email{202400130128@mail.sdu.edu.cn}

\author{Zhen Yang}
\authornote{Corresponding author.}
\affiliation{\institution{Shandong University}\city{Qingdao}\country{China}}
\email{zhenyang@sdu.edu.cn}

\author{Guanda Yan}
\affiliation{\institution{Shandong University}\city{Qingdao}\country{China}}
\email{202400271014@mail.sdu.edu.cn}

\author{Yichen Sun}
\affiliation{\institution{Henan University}\city{Zhengzhou}\country{China}}
\email{yichensun050117@gmail.com}

\author{Xuanyu Zhu}
\affiliation{\institution{Shandong University}\city{Qingdao}\country{China}}
\email{zhuxuanyu@mail.sdu.edu.cn}

\author{Zheng Shen}
\affiliation{\institution{Shandong University}\city{Qingdao}\country{China}}
\email{202400130122@mail.sdu.edu.cn}

\author{Yifei Pei}
\affiliation{\institution{Shandong University}\city{Qingdao}\country{China}}
\email{peiyifei@mail.sdu.edu.cn}

\author{Jacky Keung}
\affiliation{\institution{City University of Hong Kong}\city{Hong Kong}\country{China}}
\email{Jacky.Keung@cityu.edu.hk}

\author{Jie M. Zhang}
\affiliation{\institution{King's College London}\city{London}\country{United Kingdom}}
\email{jie.zhang@kcl.ac.uk}

\renewcommand{\shortauthors}{Ning et al.}
\begin{abstract}
Mocking is a foundational practice in unit testing, enabling developers to isolate the code under test by replacing real dependencies with simulated alternatives. Automating the generation of mock objects, stubs, and mock assertions has been a long-standing research goal. However, capture-and-replay approaches rely on recording the runtime behavior of actual dependencies. Evolutionary approaches such as StubCoder overcome this requirement but constrain generation to predefined symbol pools. Large language models (LLMs) and coding agents offer an alternative by generating mocking artifacts from source code without requiring recorded dependency executions or fixed symbol pools. Yet, no systematic study has evaluated their capabilities across all three artifact types.

To address this gap, we construct MockBench, a benchmark comprising 165 mock artifact generation tasks from real-world Java and Python projects, with validated reference mock objects, stubs, and mock assertions. We evaluate six LLMs and three coding agents in terms of executability, behavior fidelity, and interaction-fault detection capability, complemented by comparisons with existing tools and sensitivity analysis. Experimental results show that LLMs and coding agents indeed have been the state-of-the-art techniques in mock-based testing. More specifically, compared with LLMs, coding agents substantially improve executability, while these gains do not consistently translate into higher behavior fidelity or better interaction-fault detection.  
Furthermore, we analyze 660 failures from LLMs and coding agents, and identify ten error types across the three artifact categories, revealing behavioral deviations that passing tests can conceal. Our findings highlight the challenges of mock artifact generation and provide practical guidance for both developers and researchers.
\end{abstract}

\ccsdesc[500]{Software and its engineering~Software testing and debugging}
\ccsdesc[300]{Computing methodologies~Artificial intelligence}

\keywords{Mock Testing, Test Double, Large Language Model, Coding Agent}

\maketitle
\section{Introduction}
\label{Introduction}
Unit testing is a cornerstone of modern software quality assurance. In practice, a software component (e.g., a Java class or method) commonly depends on other components,
such as databases, payment gateways, file systems, or network services, thereby fulfilling its functionality \cite{mostafa2014empirical, saff2004mock}.
Testing such a component in isolation is challenging because its correctness depends on the correctness and availability of these dependencies.
Mocking addresses this challenge by replacing real dependencies with simulated substitutes, known as test doubles \cite{11025757, 10.1145/3510003.3510175},
that mimic the behavior of actual dependencies for testing purposes only. Normally, A well-formed mock-based test involves three interdependent artifacts:
(1) \textbf{Mock objects} answer the question of what to mock, i.e., which dependencies should be replaced to achieve proper test isolation. 
(2) \textbf{Stubs} answer the question of how the mock should behave. Specifically, they specify the return values, side effects,
and exception scenarios that the mock should produce when its methods are called.
(3) \textbf{Mock assertions} check whether the dependencies were used correctly, including their argument values, invocation counts, call order, etc.
Listing \ref{lst:motivating-example} illustrates a \texttt{JUnit} \cite{junit2026userguide} test case with all three mocking artifacts implemented with \texttt{Mockito} framework \cite{mockito2026}.

\noindent
\begin{minipage}{\linewidth}
\captionsetup[lstlisting]{
  font=small,
  labelfont=normalfont,
  labelsep=period,
  skip=1pt
}
\hrule height 0.4pt

\begin{lstlisting}[
  style=mockcompare,
  basicstyle=\fontsize{7}{8}\selectfont\ttfamily,
  aboveskip=1pt,
  belowskip=0pt,
  firstnumber=1
]
@Test
public void testCharge() {
  // Arrange: Prepare the dependencies and input data
  BankClient bank = mock(BankClient.class);           // Mock object
  when(bank.authorize(anyString(), anyDouble()))      // Stub call
      .thenThrow(new TimeoutException())              // First reaction
      .thenReturn(true);                              // Second reaction
  // Act: Exercise the focal method
  PaymentService svc = new PaymentService(bank);
  boolean ok = svc.charge("card-001", 150.0);
  // Assert: Verify the result and interactions
  assertTrue(ok);                                     // Oracle assertion
  verify(bank, times(2)).authorize("card-001", 150.0); // Mock assertion
}
\end{lstlisting}
\noindent
\begin{minipage}[t]{0.49\linewidth}
\begin{lstlisting}[
  style=mockcomparemethod,
  aboveskip=0pt,
  belowskip=0pt,
  firstnumber=15
]
// (a) Faulty focal method
public boolean charge(String cardId, double amount) {
  boolean ok = false;
  for (int i = 0; i < 3; i++)
    try { 
        ok = bankClient.authorize(cardId, amount);// BUG
    } catch (TimeoutException e) { /* retry */ }
  return ok;}// Oracle: PASS; Mock: FAIL (3 calls)
\end{lstlisting}
\end{minipage}\hfill
\begin{minipage}[t]{0.49\linewidth}
\begin{lstlisting}[
  style=mockcomparemethod,
  aboveskip=0pt,
  belowskip=0pt,
  firstnumber=27
]
// (b) Correct focal method
public boolean charge(String cardId, double amount) {
  for (int i = 0; i < 3; i++)
    try {
        return bankClient.authorize(cardId, amount);
    } catch (TimeoutException e) { /* retry */ }
  return false;}// Oracle: PASS; Mock: PASS (2 calls)
\end{lstlisting}
\end{minipage}
\par
\hrule height 0.4pt
\captionsetup{hypcap=false}
\captionof{lstlisting}{An example involving mock objects, stubs, and mock assertions}
\label{lst:motivating-example}
\end{minipage}
\par
The test exercises the focal method, \texttt{charge()}, whose faulty and correct implementations are shown in Lines~15-21 and Lines~27--33,
respectively, in a scenario where the payment can be authorized a charge up to three times, retrying only on a \texttt{TimeoutException}.
To isolate the focal method from the actual dependency, the test creates a mock object for \texttt{BankClient}, namely  \texttt{bank} (Line~4), and passes it to \texttt{PaymentService} as an initialization parameter (Line~10).
The stub on Lines~5--7 specifies consecutive reactions on this mock object: the first call to \texttt{authorize()} throws a \texttt{TimeoutException}, simulating a transient network fault; the second returns \texttt{true}, indicating a successful authorization, thereby testing the loop logic of the focal method with the follow-up mock assertions.
As such, the focal method, \texttt{charge()}, can be tested in isolation with mock objects and stubs. 
Continuing, the oracle assertion on Line~12 (\texttt{assertTrue(ok)}) validates the final output, while the mock assertion on Line~13 verifies that \texttt{authorize()} was invoked exactly twice on the same mock object with the specified card id and amount.
Obviously, the correct focal method immediately returns \texttt{true} at the second invocation of \texttt{authorize()}, whereas the faulty one continues the loop and makes an unnecessary third call to \texttt{authorize()}. Although they both pass the oracle assertion, the mock assertion reveals the fault of the latter.

As can be seen, systematic mocking, as in the example above, is effective for test isolation and fine-grained program verification. 
However, it is also noted that crafting all three kinds of artifacts by hand is labor-intensive and error-prone because they are tightly coupled to the specific implementations of the focal methods and their test cases,
making them even more fragile to consistent code evolution. As such, mocking artifact automation has consequently been a long-standing research goal.
Previous studies fall into three categories: (1) capture-and-replay approaches, (2) evolutionary algorithms, and (3) LLM-based testing tools. 
(1) Capture-and-replay approaches \cite{saff2004mock,saff2005automatic} generate mock artifacts by first executing the actual dependency to record its runtime interactions, then replacing the dependency with a mock object and replaying the recorded behavior. 
While straightforward, these approaches have a fundamental constraint:
they require the actual dependency to be runnable. Spadini et al.~\cite{7962389} showed that 83\% of mock-using test cases are introduced precisely when the actual dependencies are hard to set up,
flaky, or unavailable, rendering capture-and-replay inapplicable in the majority of realistic mocking scenarios.
(2) Evolutionary algorithms (e.g., StubCoder~\cite{zhu2023stubcoder}) overcame the runnability requirement by optimizing the permutations and combinations among symbols to directly produce executable mocking artifacts. Nonetheless, their symbol pools are confined to literals within the focal method and associated test file, for the sake of improving symbol reusability and avoiding search space explosion. Consequently, it cannot synthesize return values or exception types that do not already appear in this narrow scope. (3) We have also noted that some LLM-based unit test generation studies passingly produce test stubs to ensure the tests can execute, yet the quality of these stubs was not evaluated \cite{yang2026advancing,pan2025aster,nan2025test}.
We agree that LLMs or even coding agents are promising workarounds for mock-based testing, as they reason over source code semantics rather than depending on runnable dependencies or local symbol pools. Moreover, coding agents can further leverage project-wide context, build systems, and execution feedback to iteratively refine their outputs. Nonetheless, their generative quality on the end-to-end pipeline of mock objects, stubs, and assertions is unexplored in the literature.

\textbf{Empirical Study:} To fill this gap, 
we first construct \textsc{MockBench}, a benchmark comprising 165 mock artifact generation tasks from real-world Java and Python projects, with validated reference mock objects, stubs, and mock assertions.
Based on \textsc{MockBench}, we conduct a systematic study to evaluate the quality of mock artifacts generated by six LLMs and three coding agents in an end-to-end manner, where LLMs include GPT-5.5, Claude Sonnet 4.6, and Qwen3.7-Max of the commercial category, DeepSeek-v4-pro of the general category, alongside Qwen3-Coder-480B/30B of the code category. For coding agents, we experiment with \textsc{mini-SWE-agent}, \textsc{OpenCode}, and \textsc{Claude Code}. The evaluation includes executability, behavior fidelity, fault detection ability, baseline comparison, and sensitivity analysis. Moreover, we conduct a manual analysis of 660 failures made by typical LLMs and coding agents, finalizing a systematic taxonomy for future reference.

\textbf{Main Findings:} Our study yields five main findings.
(1) Compared with standalone LLMs, LLM-based coding agents substantially improve Java executability.
(2) High executability does not guarantee behavior fidelity: mock objects are easier to reconstruct than stubs and assertions, especially stub outputs and assertion arguments and constraints.
(3) Although with better executability, mock artifacts generated by coding agents do not show better behavior fidelity and fault detection ability than standalone LLMs.           
(4) Overall, LLMs and coding agents completely surpass the most relevant approach (i.e., \textsc{StubCoder}) in both executability and behavior fidelity. 
(5) Given more time budget, coding agents improve executability within a limited range, but this does not translate into better mock behavioral fidelity.
The contributions of this work are threefold:

(1) We construct \textsc{MockBench}, the first bilingual benchmark for mock artifact generation, comprising 165 Java and Python tasks with validated mocks, stubs, and assertions.

(2) We present an extensive empirical study across diverse LLMs and coding agents, assessing generated mock artifacts with test executability, behavioral fidelity, fault detection ability, baseline comparison, and sensitivity analysis.

(3) Our failure analysis reveals ten distinct error types across mock objects, stubs, and assertions masked by executable tests, offering new insights into the behavioral divergence of LLMs and coding agents and shedding light on promising directions for future improvements.
\section{Related Work}
\label{Related Work}
Mocking is a core practice for isolating dependencies in unit tests \cite{mostafa2014empirical}. Spadini et al. \cite{7962389} conducted the first large-scale empirical study on when developers choose to mock.
Ahmed et al.~\cite{ahmed2025mocking} complemented this perspective by analyzing 25,302 mocking-related questions on StackOverflow, identifying that mock configuration and framework-specific API usage remain persistent developer pain points. More recently, Hora et al. \cite{hora2026coding} analyzed over 1.2 million real-world commits on GitHub and found that coding agents add mocks more frequently than human developers, and this trend is growing. Qin \cite{qin2025mock} further revealed a divergence in mock decisions between LLMs and developers. Going beyond empirical observation, Zhu et al. \cite{zhu2020mocksniffer} proposed MockSniffer, an ML-based approach to automatically recommend which dependencies to mock, but it only outputs binary decisions rather than generating actual mock code. With the rise of LLMs for test generation, a few works have touched upon mocking. MockMill \cite{lee2026improving} extracts mocking information from existing developer tests to guide LLMs for new test generation. Notably, MocklessTester \cite{xu2026llm} takes the opposite direction by deliberately avoiding mocks, which implicitly acknowledges that handling dependencies is a known pain point for LLM-driven test generation.

Beyond deciding what to mock and generating mock objects, an equally important challenge is generating stub behavior, including the return values, side effects, and exception handling that make mocked dependencies behave realistically in tests. Early automated stub generation efforts, from capture-and-replay that records actual dependency behavior \cite{saff2004mock, saff2005automatic}, to EvoSuite's built-in stubs that target environment-dependent classes with handcrafted rules \cite{arcuri2014automated}, share a common limitation: they either require the actual dependency to be available and executable, or rely on domain-specific heuristics that cannot generalize to arbitrary dependencies. StubCoder \cite{zhu2023stubcoder} overcame these constraints by using an evolutionary algorithm guided by runtime test behavior to synthesize stub code without needing the actual dependency, and in a domain-agnostic manner.
Some automatic testing approaches also involve stub generation, such as AUTS \cite{nguyen2024automated}, Bach et al. \cite{bach2026library}, and IntUT \cite{nan2025test}, which aim to make generated tests executable; however, stub quality was not isolated as an evaluation target. Besides, Cheddadi et al. \cite{cheddadi2026google} did not propose any approach; instead, they demonstrated the use of Google Test/Google Mock to write stubs for embedded verification.

Mock assertions check whether mocked methods were invoked with expected behaviors
, which are the third essential component of mock-based testing. Zhu et al. \cite{zhu2025mockassertions} conducted the first empirical study on mock assertions, characterizing their adoption, patterns, and usage in open-source projects. Their findings provide valuable insights into how developers write mock assertions, but the study is purely descriptive without addressing automated generation. Tiwari et al. \cite{tiwari2024mimicking} proposed RICK, which monitors production execution data to generate mock-based artifacts (mock objects, stubs, and assertions) derived from production interactions to detect regression faults. However, it fundamentally depends on production deployment to collect execution traces and the Spoon metaprogramming framework for rule-based code generation, making it inapplicable during the development phase that is the focus of this study. 

In summary, prior work has made extensive contributions in mock-based testing, but most of them either focus solely on a single phase or have not yet touched upon LLMs and coding agents for a systematic evaluation. Compared to traditional methods, LLMs and agents offer greater flexibility in generation and hold significant potential for further enhancing the applicability of automated mock testing. This paper, therefore, conducts a systematic study to address this gap.


\section{New Benchmark: Mockbench}\label{sec:mockbench}
\subsection{Task Definition}\label{sec:task}
As illustrated in Listing~1, the unit tests considered in \textsc{MockBench}
are organized according to the Arrange--Act--Assert (AAA) pattern~\cite{yu2019comprehending}.
The Arrange phase creates and configures mock objects, including their stub
calls; the Act phase invokes the focal method, i.e., Method Under Test (MUT);
and the Assert phase checks the test outcome through oracle assertions and,
when applicable, mock assertions that verify expected interactions with mocks.

To shrink the experimental cost, we process each complete test suite as one generation unit. Developer-written test suites $T^G$ are used as ground truth.
Within $T^G$, the three types of mock artifacts: mock-object declarations,
stub calls, and mock assertions, are denoted by
$F_{\mathrm{mock}}^G$, $F_{\mathrm{stub}}^G$, and
$F_{\mathrm{ver}}^G$, respectively.
By removing these three mock artifacts from $T^G$, we obtain
the masked test suite $\widetilde{T}$.
Given $\widetilde{T}$ and the MUT, an LLM/coding agent $M$, generates the corresponding candidate mock artifacts:
$(F_{\mathrm{mock}}^C, \, F_{\mathrm{stub}}^C, \, F_{\mathrm{ver}}^C) = M(\widetilde{T}, \mathit{MUT})$.
Afterwards, we leverage predefined scripts to automatically inject the three candidate artifacts into $\widetilde{T}$, forming the candidate test suite $T^C$ for subsequent evaluation:$T^C = \widetilde{T} \triangleleft (F_{\mathrm{mock}}^C, \, F_{\mathrm{stub}}^C, \, F_{\mathrm{ver}}^C)$.

\subsection{MockBench Construction}\label{sec:benchmark}
This section introduces \textsc{MockBench}, a bilingual benchmark for evaluating mock artifact generation across real-world Java and Python projects. It provides ground-truth mock objects, stubs, and assertions for both ecosystems. We describe the benchmark construction and data statistics below.

\subsubsection{Benchmark Design}~\\
\textbf{Java Portion}
For the Java portion, we build upon StubCoder~\cite{zhu2023stubcoder}, which provides 59 real-world test cases across 13 projects with genuine mock configurations, production code, and compilation dependencies.
However, not all external dependencies are mocked, and the existing mock assertions do not always fully specify the expected dependency interactions.
Therefore, further refinement is needed for our evaluation.

Specifically, following a pervasive mocking practice \cite{spadini2019mock,arcuri2017mocking}, we first select MUT-dependent objects that come from third-party libraries or outside the class of the MUT for mocking.
Next, for each selected dependency in the MUT, we progressively add mock objects, stubs, and mock assertions if they are missing in the original test. 
Our modification does not change the production code, focal calls, test inputs, or fixtures.
We also augmented the suites with oracle assertions to verify the observable outcomes of stubbed behaviors. 
In total, 53 of the 59 Java test suites received additional checks: oracle assertions increased from 117 to 201, while mock assertions rose from 61 to 126. 
Consequently, all 59 suites now enforce explicit interaction verifications, up from only 30 in the original baseline.
We implement tests using \texttt{JUnit}~\cite{junit2026userguide} alongside \texttt{Mockito}~\cite{mockito2026}, managing dependencies through \texttt{Maven}~\cite{apacheMaven} or \texttt{Gradle}~\cite{gradle}.

\textbf{Python Portion}
Existing Python testing benchmarks focus on test generation and coverage assessment \cite{wang-etal-2025-testeval,jain2024testgeneval}, or provide reproducible real-world bugs for testing and debugging studies \cite{widyasari2020bugsinpy}. These evaluation objectives differ from our focus on mocking artifact generation. We therefore construct a dedicated Python benchmark tailored to this task.

Specifically, we curate 106 MUTs from 12 open-source projects across 6 application domains with rich dependencies and high popularity (Table~\ref{tab:benchmark}). 
Nonetheless, most MUTs in the wild are not equipped with sufficient test cases. Therefore, we manually construct a reference test suite for each Python MUT to capture multiple representative dependency-interaction scenarios.
Each suite uses parameterized tests to instantiate these scenarios by varying method inputs and stubbed dependency responses while specifying the corresponding expected interactions, without modifying production code.
Furthermore, each reference suite is required to achieve at least 90\% line coverage of its MUT to ensure sufficient exercise of the focal logic.
The Python portion uses \texttt{pytest} ~\cite{pytest2026}as the testing framework and \texttt{pytest-mock}~\cite{pytestmock2026} for mock-object construction and interaction verification. 
\begin{table}[t]
\footnotesize
\setlength{\abovecaptionskip}{0pt}
\caption{Statistical Overview of the MockBench}
\label{tab:benchmark}
\centering
\begingroup
\scriptsize
\setlength{\tabcolsep}{0pt}
\renewcommand{\arraystretch}{0.76}
\setlength{\aboverulesep}{0.5pt}
\setlength{\belowrulesep}{0.5pt}
\begin{tabular*}{\linewidth}{@{\extracolsep{\fill}} l l r rr rrr S[table-format=3.1]@{\extracolsep{0pt}\,$\pm$\,}S[table-format=2.1] @{}}
\toprule
\textbf{Domain} & \textbf{Project} & \textbf{Stars} & \textbf{\#Suite} & \textbf{\#Case} & \textbf{\#Obj} & \textbf{\#Stub} & \textbf{\#Assert} & \multicolumn{2}{c}{\textbf{LOC/Suite}} \\
\midrule
\multirow{4}{*}{Data}
    & druid             & 14.1k &  2 &  2 &  7 &  5 &  5 &  94.5 & 36.1 \\
    & hadoop            & 15.7k &  5 &  5 & 11 & 19 & 17 & 122.0 & 32.9 \\
    & mybatis-3         & 20.4k & 12 & 12 & 37 & 12 & 24 &  44.6 &  2.2 \\
    & neo4j             & 17.3k &  4 &  4 & 35 & 27 & 13 & 109.0 & 30.7 \\
\multirow{3}{*}{Dist}
    & dubbo             & 41.6k & 13 & 13 & 29 & 48 & 17 &  67.3 & 12.4 \\
    & zookeeper         & 12.8k &  1 &  1 &  2 &  4 &  1 & \multicolumn{2}{c}{131.0} \\
    & grpc-java         & 12.1k &  3 &  3 &  5 &  3 &  3 &  57.7 &  8.4 \\
\multirow{4}{*}{Ops}
    & apollo            & 29.8k &  8 &  8 & 43 & 35 & 20 &  89.9 & 45.4 \\
    & spring-boot-admin & 12.9k &  2 &  2 &  2 &  2 &  2 &  63.5 &  0.7 \\
    & jib               & 14.5k &  2 &  2 &  9 & 16 &  4 & 102.0 & 18.4 \\
    & zipkin            & 17.5k &  5 &  5 & 18 & 18 & 18 &  55.8 &  1.1 \\
\multirow{2}{*}{Web}
    & zuul              & 14.1k &  1 &  1 &  2 &  1 &  1 & \multicolumn{2}{c}{68.0} \\
    & spring-boot       & 81.5k &  1 &  1 &  1 &  1 &  1 & \multicolumn{2}{c}{45.0} \\
\midrule
\multicolumn{2}{l}{\textbf{Total (Java Portion)}} & & \textbf{59} & \textbf{59} & \textbf{201} & \textbf{191} & \textbf{126} & \multicolumn{2}{c}{$\mathbf{74.4 \pm 33.2}$} \\
\midrule
\multirow{2}{*}{Web}
    & flask                  &  71.7k & 26 & 110 & 80 & 93 & 75 &  87.3 & 20.1 \\
    & redash                 &  28.7k &  6 &  20 & 16 & 19 & 25 & 109.5 &  9.5 \\
CLI & click                  &  17.6k & 11 &  61 & 31 & 31 & 33 & 126.5 & 17.1 \\
\multirow{4}{*}{ML}
    & langflow               & 150.0k & 11 &  35 & 28 & 28 & 31 &  91.4 & 23.4 \\
    & stable-diffusion-webui & 163.9k &  7 &  33 & 16 & 16 & 17 & 165.4 & 33.2 \\
    & verl                   &  22.1k &  7 &  38 & 25 & 27 & 37 &  99.6 & 44.9 \\
    & ultralytics            &  58.8k &  5 &  18 & 14 & 16 & 21 &  80.0 &  0.0 \\
\multirow{2}{*}{Ops}
    & ansible                &  69.0k &  6 &  39 &  8 & 13 & 11 & 156.8 & 63.1 \\
    & crawl4ai               &  69.4k &  8 &  33 & 26 & 23 & 39 & 126.2 & 14.7 \\
\multirow{2}{*}{Finance}
    & hummingbot             &  19.0k &  7 &  27 & 15 & 24 & 15 & 129.0 & 29.6 \\
    & tqsdk-python           &   4.8k &  7 &  25 & 18 & 19 & 29 &  64.7 & 10.1 \\
NLP & spaCy                  &  33.7k &  5 &  22 & 20 & 21 & 33 & 217.0 &  0.0 \\
\midrule
\multicolumn{2}{l}{\textbf{Total  (Python Portion)}} & & \textbf{106} & \textbf{461} & \textbf{297} & \textbf{330} & \textbf{366} & \multicolumn{2}{c}{$\mathbf{112.9 \pm 43.7}$} \\
\bottomrule
\end{tabular*}
\endgroup
\end{table}

\subsubsection{Construction Procedure}
Following the aforementioned design, seven developers participated in constructing \textsc{MockBench}, with one serving as the lead reviewer and the remaining six responsible for benchmark construction.
Each developer independently processed the assigned subjects—refining reference tests for Java and constructing parameterized reference suites for Python according to the established guidelines.
The lead reviewer then examined all completed items to verify dependency selection, scenario design, and mock-artifact completeness, resolving any disputed or borderline cases.
Finally, all reviewed tests were validated via execution: Java tests had to compile and pass, whereas Python suites were required to pass all parameterized test cases and meet the predefined MUT line-coverage threshold.


\section{Experimental Design}\label{design}


This study uses \textsc{MockBench} to evaluate LLMs and coding agents on mock artifact generation with the following five Research Questions (RQs):

\textbf{RQ1 (Executability): \emph{How do LLMs and coding agents perform in generating executable mock artifacts?}}
Generating mock-object declarations, stub calls, and mock assertions together requires coordinating how dependencies are replaced, how they respond, and which interactions are checked.
We first examine whether LLMs and coding agents can deliver $T^C$ with auto-generated mocking artifacts that make the MUT pass all the tests.

\textbf{RQ2 (Behavior Fidelity): \emph{To what extent do $T^C$ faithfully reproduce the mock behavior of $T^G$ ?}}
To examine the generation quality of $T^C$ meticulously and systematically, we design scripts that automatically localize behavioral omissions and deviations in mock objects, stubs, and mock assertions relative to $T^G$.

\textbf{RQ3 (Interaction-Fault Detection Capability):\emph{To what extent do  $T^C$ preserve the fault detection capability encoded by developer-written references $T^G$?}}
An executable $T^C$ does not necessarily mean it preserves the $T^G$'s sensitivity to incorrect interactions between dependencies and the MUT, namely interaction faults. Such faults are responsible for mock assertions to examine, which is different from I/O faults revealed by oracle assertions.
As such, this RQ intends to use mutant analysis~\cite{jia2011mutation,just2014mutants,andrews2005mutation,papadakis2019mutation} to uncover the above question, providing a dynamic quality dimension complementary to the behavior fidelity studied in RQ2.


\textbf{RQ4 (Baseline Comparison): \emph{How do LLMs and coding agents perform relative to dedicated mock generation tools?}}
Among existing mock-related techniques, StubCoder is the only comparable baseline: similar to LLMs and coding agents, it synthesizes stub behaviors from code and test context without executing real dependencies. Nevertheless, its native scope is limited to stub generation, whereas we tackle the holistic generation of mock objects, stubs, and mock assertions.


\textbf{RQ5 (Resource Sensitivity): \emph{How does the operational time budget influence the effectiveness and workflow of coding agents?}}
Coding agents can inspect projects, revise candidates, and perform validation during generation, while the time budget constrains the opportunities for these activities.
We examine how final generation quality and workflows vary across budgets to understand the gains from additional time and the associated resource consumption.

\subsection{Studied subjects}\label{sec:subjects}

\begingroup
\setlength{\intextsep}{6pt}
\renewcommand{\arraystretch}{0.9} 
\begin{table}[!htb]
\scriptsize
\setlength{\abovecaptionskip}{0pt}
\caption{Studied LLMs in this paper}
\label{tab:llms}
\centering
\footnotesize
\setlength{\tabcolsep}{3pt}
\renewcommand{\arraystretch}{0.8}
\setlength{\aboverulesep}{0.5pt}
\setlength{\belowrulesep}{0.5pt}
\begin{tabular*}{\linewidth}{@{\extracolsep{\fill}}l l c c c c c@{}}
\toprule
Category & Model & Time & Size & MoE/RoPE & Training Base (tokens) & Context (tokens) \\
\midrule
\multirow{3}{*}{Commercial}
    & GPT-5.5~\cite{openai2025gpt55}                & 2026-04 & ---       & Y/---   & ---  & 1{,}050K/128K \\
    & Claude Sonnet 4.6~\cite{anthropic2025sonnet46} & 2026-02 & ---       & ---/--- & ---  & 1{,}000K/64K  \\
    & Qwen3.7-Max~\cite{qwen2025qwen37max}          & 2026-05 & ---       & Y/---   & ---  & 1{,}000K/65.5K \\
\midrule
General
    & DeepSeek-v4-pro~\cite{deepseek2025v4pro}      & 2026-04 & 1.6T/49B  & Y/Y     & ---  & 1{,}000K/64K \\
\midrule
\multirow{2}{*}{Code}
    & Qwen3-Coder-480B~\cite{qwen2025qwen3coder}    & 2025-07 & 480B/35B  & Y/Y     & 7.5T & 256K/65K \\
    & Qwen3-Coder-30B~\cite{qwen2025qwen3coder}     & 2025-07 & 30B/3.3B  & Y/Y     & 7.5T & 256K/65K \\
\bottomrule
\end{tabular*}
\par
\begin{minipage}{\linewidth}
\scriptsize\raggedright
\textit{Note.} All models are instruction-following models. In the following sections, ``DeepSeek'' and ``Claude'' refer to DeepSeek-v4-pro and Claude Sonnet~4.6; ``Qwen-480B'' and ``Qwen-30B'' denote the two Qwen3-Coder variants.
\end{minipage}
\end{table}
\endgroup

\noindent\textbf{Studied LLMs}\label{sec:llm_setting}
Table~\ref{tab:llms} lists the six LLMs studied in this work with their categories, model types, release times(Time), parameter sizes(Size), architectural characteristics(MOE/RoPE),
pretraining token numbers(Training Base), and context-window limits(Context).
To extensively explore the performance of recent LLMs on mock artifact generation, we select LLMs from commercial, general-purpose,
and code-specialized categories.
Specifically, the commercial models include GPT-5.5~\cite{openai2025gpt55},Claude Sonnet 4.6~\cite{anthropic2025sonnet46}, and Qwen3.7-Max~\cite{qwen2025qwen37max};
the general-purpose model is DeepSeek-v4-pro~\cite{deepseek2025v4pro};
and the code-specialized models include Qwen3-Coder-480B and Qwen3-Coder-30B~\cite{qwen2025qwen3coder}.
Furthermore, for LLMs from the same family, we select two Qwen3-Coder variants of different sizes, i.e., Qwen3-Coder-30B and Qwen3-Coder-480B,
to study the performance difference across model scales.

\noindent\textbf{Studied Coding Agents}\label{sec:agent_setting}
To cover different agent designs, we select three coding agents:
\textsc{mini-SWE-agent}~\cite{yang2024sweagent}, \textsc{OpenCode}~\cite{opencode2025}, and \textsc{Claude Code}~\cite{anthropic2025claudecode}.
\textsc{mini-SWE-agent} represents a lightweight agent scaffold with a minimal command-line interaction design,
whereas \textsc{OpenCode} represents a full-featured open-source coding agent with repository-level interaction.
\textsc{Claude Code} further represents a vendor-developed, full-featured coding agent designed for agentic software development workflows.
To control for the influence of the underlying LLM, we configure all three coding agents with DeepSeek as the common backbone,
enabling comparison across different agent workflows under the same model configuration.


%
%
\subsection{Evaluation Protocol and Metrics}\label{sec:evaluation_metrics}

This section introduces the evaluation procedure and metrics for each RQ.

\textbf{RQ1 (Executability)}
As mentioned in Section \ref{sec:task}, for every masked test suite $\widetilde{T}$, we generate three mock artifacts holistically, including mock objects, stubs, and mock assertions, and reassemble them into $\widetilde{T}$, obtaining a candidate test suite $T^C$ for evaluation.
RQ1 measures executability through Compile pass rate (CPR), Suite pass rate (SPR), and
Test case pass rate (TPR), where CPR is the fraction of $T^C$ that compile successfully, SPR is the fraction of $T^C$ in which all test cases pass, and TPR measures the proportion of passing test cases within $T^C$. Note that the Java portion of \textsc{MockBench}, inherited from \textsc{StubCoder}~\cite{zhu2023stubcoder}, contains one test case per suite, so SPR and TPR coincide on Java.
The two metrics diverge only on the Python benchmark, where a test suite contains multiple parameterised test cases.

\textbf{RQ2 (Behavior Fidelity)}
To examine the faithfulness of $T^C$ towards $T^G$, we parse the AST of both tests to structurally extract their interaction records of mock objects, stubs, and mock assertions. In general, each statement of \textbf{mock object declaration} can be recorded as a pair
$\langle \mathit{creator}, \mathit{target} \rangle$, where $\mathit{creator}$ is the invocation of the mock creation API, while $\mathit{target}$ is the class being mocked.
Besides, each statement of a \textbf{stub call} can be recorded as a five-field tuple
$\langle \mathit{recv}, \mathit{meth}, \mathit{args},
\mathit{kind}, \mathit{val} \rangle$,  where $\mathit{recv}$ is the mock object instantiated from a $\mathit{target}$, $\mathit{meth}$ is an instance method of $\mathit{recv}$ invoked by the MUT, $\mathit{args}$ are the arguments
passed to $\mathit{meth}$, $\mathit{kind}$ is the stub response type (e.g., return value, thrown exception), and $\mathit{val}$ is the corresponding output value.
Finally, each \textbf{mock assertion} is analogously recorded as a four-tuple $\langle \mathit{recv}, \mathit{meth}, \mathit{args},
\mathit{constr} \rangle$, where $\mathit{constr}$ captures the verification constraint imposed on the expected interaction.
Taking Listing~\ref{lst:motivating-example} as an example, the mock-object declaration at Line~4 yields a mock record:
$\mathit{creator}$ = \texttt{mock()} and
$\mathit{target}$ = \texttt{BankClient}.
The stub at Lines~5--7 specifies two consecutive responses to
\texttt{bank.authorize(anyString(), anyDouble())}:
the first invocation throws a \texttt{TimeoutException}, and the second returns \texttt{true}.
Taking the return response at Line~7 as an example, the five fields are
$\mathit{recv}$ = \texttt{bank},
$\mathit{meth}$ = \texttt{authorize},
$\mathit{args}$ = \texttt{(anyString(), anyDouble())},
$\mathit{kind}$ = \texttt{return}, and
$\mathit{val}$ = \texttt{true}.
The mock assertion at Line~15 yields a record with four fields:
$\mathit{recv}$ = \texttt{bank},
$\mathit{meth}$ = \texttt{authorize},
$\mathit{args}$ = \texttt{("card-001", 150.0)}, and
$\mathit{constr}$ = \texttt{times(2)}.
Since there are different APIs with equivalent functionalities (e.g., \texttt{thenReturn} and \texttt{doReturn}), we normalize them to the same name in records, thereby facilitating a more accurate evaluation. In addition, objects in $\mathit{recv}$, $\mathit{args}$, and $\mathit{val}$ are traced to their underlying classes, so that naming differences (e.g., \texttt{repo} and \texttt{repository}) do not cause spurious mismatches. 
After fetching the records for both $T^G$ and $T^C$, we progressively extract fields for evaluation in a coarse\mbox{-}to\mbox{-}fine manner. 

\textbf{(1) Coarse-grained evaluation.} To assess the full behavior fidelity, we extract all corresponding fields from both $T^G$ and $T^C$ for assessment.
\textbf{(2) Fine-grained evaluation.} Considering that stub calls and mock assertions are more complicated with many diverse fields in comparison with mock object declaration, we dive deeper into these two artifacts with a series of fine-grained metrics for further assessment. For \textbf{stub calls}, we partition the evaluation into four aspects: (1) At the method aspect, we extract both the $\langle \mathit{recv}, \mathit{meth} \rangle$ fields to assess whether the stub invokes the correct method. Both fields are included for evaluation because only considering the correctness of $\mathit{meth}$ is meaningless if the $\mathit{recv}$ is incorrect.
(2) At the argument aspect, $\langle \mathit{recv}, \mathit{meth}, \mathit{args} \rangle$ fields are extracted in a cascaded manner for a similar reason to assess whether the stub passes the correct arguments. (3) At the response aspect, we extract $\langle \mathit{recv}, \mathit{meth},  \mathit{kind} \rangle$ fields to assess whether the stub selects the correct response type.
(4) At the output aspect, we extract  $\langle \mathit{recv}, \mathit{meth}, \mathit{val}  \rangle$ fields to assess whether a stub reconstructs the intended return value or exception behavior.
Towards \textbf{mock assertions}, the evaluation is divided into three aspects, where the method and argument aspects use the same fields as those in the stub evaluation.
Apart from the above, a mock assertion should also verify the invocation frequency of a method. Therefore, we propose an extra evaluation aspect, namely constraint aspect,
which requires  ($\langle \mathit{recv}, \mathit{meth},  \mathit{constr} \rangle$)
to participate in this metric measurement.

Following the above procedure, we evaluate each mock-artifact type
separately under the corresponding record representation.
Let $C=\{c_1,\ldots,c_m\}$ and $G=\{g_1,\ldots,g_n\}$
denote the sets of normalized records extracted from the full sets $T^C$ and $T^G$,
respectively.
Each $c_i$ or $g_j$ represents one record containing all fields for
coarse-grained evaluation or the selected fields for a fine-grained aspect.
We use $[c_i==g_i]$ to denote 1 if all included fields are identical, and 0 otherwise. 
We then compute precision($P=\frac{\sum_{i=1}^{m}[c_i==g_i]}{|C|}$), recall($R=\frac{\sum_{i=1}^{m}[c_i==g_i]}{|G|}$), and F1-score($F1=\frac{2PR}{P+R}$).

\textbf{RQ3 (Interaction-Fault Detection Capability)}
Following prior mutation-based
evaluations~\cite{tiwari2024mimicking,zhu2023stubcoder}, we compare the mutants detected by $T^C$ and $T^G$. 
However, general-purpose mutation tools such as PIT~\cite{coles2016pit}
and Cosmic Ray~\cite{cosmicray} do not specifically target the collaborator interactions checked by mock assertions~\cite{zhang2025min}. Our preliminary experiments also found that mutants produced by these tools were often caught by oracle assertions before reaching the interactions checked by mock assertions, making it difficult to distinguish the interaction-fault detection capabilities of $T^G$ and $T^C$.
Mock assertions check interactions between collaborators and the MUT in terms of whether calls occur, how often they occur, the arguments they receive, whether unintended calls occur, and their execution order~\cite{zhu2025mockassertions,freeman2004mockroles}. Accordingly, we design five first-order mutation operators, each introducing a single fault into one of these interaction dimensions~\cite{jia2009higherorder}.
(1)~\textbf{DuplicateCall}: duplicates a call, e.g., \texttt{authorize()} $\to$ \texttt{authorize(); authorize()};
(2)~\textbf{AlterArgument}: modifies an argument, e.g., \texttt{cache.get(key)} $\to$ \texttt{cache.get("MUTATED")};
(3)~\textbf{DropCall}: removes a call, e.g., \texttt{audit.record(id)} $\to$ $\varnothing$;
(4)~\textbf{InsertForbiddenCall}: inserts an absent call, e.g., $\varnothing$ $\to$ \texttt{dao.delete(id)}; and
(5)~\textbf{AlterOrder}: swaps two ordered calls, e.g., \texttt{authorize(); settle()} $\to$ \texttt{settle(); authorize()}.





Afterwards, we use DeepSeek-v4-pro to perform these transformations with specific mutation prompts on each of the MUTs in MockBench.
We retain only compilable mutants that pass all oracle assertions but fail at least one mock assertion in $T^G$, thereby isolating faults detected through interaction verification.
Within this retained mutant set, let $K_G$ and $K_C$ denote the sets of mutants killed by $T^G$ and the corresponding $T^C$, respectively.
Following \cite{tiwari2024mimicking,zhu2023stubcoder},
we report the kill-set Jaccard similarity between $K_G$ and $K_C$ to measure the fault detection ability of $T^C$.
Recall that all retained mutants are killed by mock assertions of $T^G$; thus, $K_G$ equals the full retained mutant set, and $K_C \subseteq K_G$.
The Jaccard similarity therefore simplifies to:
$J = |K_C|/|K_G|$, which measures the kill-set overlap between $T^G$ and $T^C$. 

\textbf{RQ4 (Baseline Comparison)}
StubCoder is a stub generation tool based on the evolutionary algorithm, but it applies only to Java. To align the comparison with StubCoder's setting, we experiment on the Java portion of MockBench and mask only the stub region while retaining the mock objects and mock
assertions from $T^G$ as the model input \cite{zhu2023stubcoder}.
To control the experimental scale, we select DeepSeek-v4-pro, a representative LLM, and three coding agents to make the comparison with StubCoder.
As for evaluation metrics, we adopt TPR to assess the executability of auto-generated stubs within $T^C$, and leverage the F1-scores of stub reconstruction's four dimensions (i.e., Method, Argument, Response, and Output) to examine its behavior fidelity.

\textbf{RQ5 (Resource Sensitivity)}
Following prior agent evaluations that examine performance under different resource budgets~\cite{ni2026gittaskbench,chan2025mlebench,starace2025paperbench}, we evaluate the three coding agents on MockBench under per-suite wall-clock budgets of
60\,s, 300\,s, and 600\,s.
All agents use DeepSeek-v4-pro as their backbone and follow the execution and candidate-retention rules described in Section~\ref{sec:implementation}.
We report TPR, F1-score of coarse-grained behavior fidelity for mock objects, stubs, and mock assertions, and resource consumption measured by agent turns and token usage.
\subsection{Implementation Details}\label{sec:implementation}
Given the MUT and the corresponding $\widetilde{T}$, each LLM is prompted to generate the mock artifacts.
We access all evaluated LLMs through their official APIs.
For standalone generation, we set the temperature to 0 with $n=1$ and evaluate the single returned candidate.
To account for residual nondeterminism, we repeat each experiment three times under identical configurations, reporting the mean and standard deviation across runs.

For agent-based generation, each benchmark evaluation runs within an isolated Docker container with a dedicated workspace allocated per generation task.
To simulate scenarios where concrete dependency implementations are inaccessible, we deliberately strip the corresponding implementations from the workspace.
Within the workspace, agents interact with the repository via a curated set of tools: inspecting files (\texttt{list\_files}, \texttt{read\_file}), executing build and testing workflows (\texttt{run\_shell}), and emitting generated mocks (\texttt{write\_file}).
In the main experiments, each task has a 300\,s wall-clock limit and terminates upon an explicit agent exit or budget exhaustion.
We checkpoint a candidate only after it passes all tests in $T^C$. 
Upon timeout, we return the latest checkpoint to avoid partially modified code, or \texttt{None} if no checkpoint exists.
Containers run without network access and are destroyed upon completion.

\section{Experimental Results}\label{sec:results}

\subsection{RQ1: Executability}\label{sec:rq1}

\begin{table}[!htbp]
\centering
\setlength{\abovecaptionskip}{0pt}
\caption{RQ1: Executability results (\%).}
\label{tab:rq1}
\scriptsize
\setlength{\tabcolsep}{1.8pt}
\renewcommand{\arraystretch}{0.76}
\setlength{\aboverulesep}{0.5pt}
\setlength{\belowrulesep}{0.5pt}
\begin{tabular*}{\linewidth}{@{\extracolsep{\fill}}l cc ccc@{}}
\toprule
& \multicolumn{2}{c}{\textit{Java}}
& \multicolumn{3}{c}{\textit{Python}} \\
\cmidrule(lr){2-3}
\cmidrule(lr){4-6}
Subject & CPR & TPR & CPR & SPR & TPR \\
\midrule

Claude
  & \bestcell{92.1}\sd{0.8}
  & 68.6\sd{2.5}
  & 97.5\sd{1.29}
  & 85.2\sd{2.23}
  & 90.7\sd{0.9} \\

GPT-5.5
  & 88.1\sd{1.71}
  & 71.2\sd{2.8}
  & 98.13\sd{1.61}
  & 91.4\sd{1.37}
  & \bestcell{96.1}\sd{0.8} \\

Qwen3.7-Max
  & 87.3\sd{0.8}
  & 69.5\sd{1.7}
  & \bestcell{98.33}\sd{0.72}
  & 88.9\sd{1.06}
  & 93.6\sd{0.7} \\

DeepSeek
  & 59.3\sd{6.8}
  & 44.1\sd{8.5}
  & 95.83\sd{0.73}
  & 82.7\sd{1.2}
  & 86.5\sd{1.1} \\

Qwen-480B
  & 77.1\sd{0.8}
  & 48.3\sd{2.5}
  & 96.15\sd{2.36}
  & 74.1\sd{2.36}
  & 81.6\sd{1.3} \\

Qwen-30B
  & 63.6\sd{2.5}
  & 35.6\sd{1.33}
  & 97.54\sd{1.33}
  & 70.4\sd{2.17}
  & 74.8\sd{1.4} \\

\addlinespace[1pt]
\textit{LLM Avg.}
& 77.92\sd{13.75} & 56.22\sd{15.42}
& 97.25\sd{1.03} & 82.12\sd{8.29} & 87.22\sd{7.98} \\

\specialrule{0.05em}{1.5pt}{1.5pt}

Claude Code
  & 71.19\sd{6.11}
  & 71.19\sd{6.11}
  & 79.56\sd{4.84}
  & 79.56\sd{4.84}
  & 79.18\sd{5.78} \\

mini-SWE-agent
  & 89.21\sd{2.68}
  & \bestcell{88.65}\sd{2.66}
  & 95.91\sd{1.44}
  & \bestcell{95.91}\sd{1.44}
  & 95.95\sd{2.17} \\

OpenCode
  & 83.62\sd{7.64}
  & 83.05\sd{6.78}
  & 84.28\sd{10.89}
  & 84.28\sd{10.89}
  & 82.79\sd{11.25} \\

\addlinespace[1pt]
\textit{Agent Avg.}
& 81.34\sd{9.22} & 80.96\sd{8.92}
& 86.58\sd{8.41} & 86.58\sd{8.41} & 85.97\sd{8.83} \\

\bottomrule
\end{tabular*}
\textit{Note.} Green shading marks per-language column maxima across LLMs and agents, here and in subsequent tables.
\end{table}
\textbf{Comparison across PLs.}
LLM-crafted $T^C$s achieve consistently higher executability in Python than in Java.
Specifically, the average TPR reaches 87.2\% on Python compared to only 56.2\% on Java, a gap of 31.0 pp.
A similar trend appears in terms of CPR, which, on average, is only 78.0\% on Java, compared to 97.2\% on Python.
The reason is that Java imposes stricter compile-time requirements: models must correctly specify types, imports, class and method names, method signatures, and object construction, where an error in any of these elements can prevent compilation.
\begin{findingbox}
\textbf{Finding 1:}
All evaluated LLMs consistently perform worse in Java than in Python, from initial compilation to full test execution. The gap is driven by Java's strict type constraints and verbose symbol requirements.
\end{findingbox}
\textbf{Comparison among LLMs.}
As shown in Table~\ref{tab:rq1}, the three commercial LLMs (i.e., GPT-5.5, Claude, and Qwen3.7-Max) lead in TPR across both Java and Python.
These models benefit from strong general reasoning and long-context capabilities, allowing them to follow complex prompt instructions and handle mock setup constraints more reliably.
When comparing general and code-specialized models, general LLMs hold a clear edge.
Specifically, Qwen3.7-Max consistently outperforms both Qwen3-Coder variants across both languages.
Compared to Qwen3-480B, Qwen3.7-Max achieves a TPR that is 21.2 pp higher on Java and 12.0 pp higher on Python.
This gap suggests that synthesizing mock artifacts requires more than familiar code patterns; it demands deep reasoning about method execution paths and dependency behaviors, where massive general-purpose models excel.
Within the same family, model scale remains a primary performance driver.
Qwen3-480B substantially beats Qwen3-30B on both languages, lifting TPR by 12.7 pp on Java and 6.8 pp on Python.
The larger parameter budget especially benefits Java, where strict static typing and verbose framework APIs require precise symbol resolution and syntax accuracy.
\begin{findingbox}
\textbf{Finding 2:}
Compared with open-source LLMs, commercial LLMs exhibit consistently superior performance on test executability.
Besides, code LLMs do not necessarily outperform general LLMs, and smaller LLMs perform worse within the same family, given other factors the same.
\end{findingbox}
\textbf{Executability across Coding Agents.}
Coding agents greatly improve test executability, especially on the harder Java benchmark.
With the shared DeepSeek backbone, agent TPR on Java increases from 44.1\% under standalone inference to 71.19--88.65\%, an absolute gain of 27.1--44.6 percentage points.
On Python, standalone DeepSeek already achieves a strong 86.5\% TPR, leaving less room for improvement; here, agents achieve 79.18--95.95\% TPR.

The remaining failures happen because agents exhaust their time budget, rather than generating code that fails to run.
Once agents deliver a candidate, downstream compilation and test errors almost disappear: all three agents reach 100\% CPR across both languages, with conditional TPR exceeding 99.3\% on Java and reaching 100\% on Python.
Thus, the main barrier to executability is whether agents can finish code generation within their budget.
\begin{findingbox}
\textbf{Finding 3:}
Coding agents substantially boost test executability, raising Java TPR by 27.1--44.6 pp over standalone models.
Remaining failures stem from timeouts rather than test crashes: once agents deliver code, they achieve near-perfect pass rates across both languages.
\end{findingbox}
\subsection{RQ2: Behavior Fidelity}\label{sec:rq2}

\begin{table}[!htb]
\setlength{\abovecaptionskip}{0pt}
\caption{RQ2: coarse-grained behavior fidelity for mock objects, stubs, and mock assertions. }
\label{tab:rq2_component}
\centering
\scriptsize
\setlength{\tabcolsep}{1.45pt}
\renewcommand{\arraystretch}{0.80}
\setlength{\aboverulesep}{0.5pt}
\setlength{\belowrulesep}{0.5pt}
\resizebox{\columnwidth}{!}{%
\begin{tabular}{@{}l@{\hspace{3pt}}l *{3}{ccc}@{}}
\toprule
& & \multicolumn{3}{c}{Mock object}
& \multicolumn{3}{c}{Stub}
& \multicolumn{3}{c}{Mock assertion} \\
\cmidrule(lr){3-5}
\cmidrule(lr){6-8}
\cmidrule(lr){9-11}
& Subject
& P & R & F1
& P & R & F1
& P
& R
& F1 \\
\midrule

\multirow{11}{*}[-3pt]{\rotatebox{90}{\textbf{Java}}} & GPT-5.5
& 91.25\sd{2.45} & 76.28\sd{4.83} & 83.09\sd{3.88} & 63.46\sd{6.22} & 65.41\sd{2.61} & 64.39\sd{4.48} & 47.90\sd{3.56} & \bestcell{65.95\sd{2.58}} & 55.49\sd{3.30} \\

& Claude
& 93.70\sd{0.67} & \bestcell{77.53\sd{5.72}} & 84.79\sd{3.16} & 57.25\sd{2.78} & 61.00\sd{3.37} & 59.06\sd{3.06} & 63.17\sd{10.26} & 58.98\sd{0.78} & 60.83\sd{5.21} \\

& Qwen3.7-Max
& 93.96\sd{1.74} & 72.04\sd{0.43} & 81.55\sd{0.38} & 64.12\sd{4.04} & 54.92\sd{4.26} & 59.16\sd{4.19} & 69.57\sd{4.10} & 52.46\sd{1.84} & 59.81\sd{2.71} \\

& DeepSeek
& 94.41\sd{3.41} & 56.44\sd{16.04} & 70.13\sd{13.59} & 73.81\sd{5.61} & 63.25\sd{0.55} & 68.08\sd{2.71} & \bestcell{83.29\sd{5.18}} & 61.15\sd{10.04} & \bestcell{70.09\sd{4.83}} \\

& Qwen-480B
& 96.30\sd{0.00} & 65.41\sd{0.58} & 77.90\sd{0.41} & 73.73\sd{0.06} & \bestcell{77.68\sd{0.14}} & 75.65\sd{0.03} & 81.57\sd{0.69} & 55.61\sd{6.74} & 66.03\sd{5.00} \\

& Qwen-30B
& 97.47\sd{0.05} & 52.74\sd{0.97} & 68.44\sd{0.83} & 53.75\sd{5.30} & 43.41\sd{3.67} & 48.03\sd{4.36} & 71.67\sd{2.36} & 47.78\sd{1.57} & 57.33\sd{1.89} \\

& \textit{LLM Avg.}
& 94.52\sd{2.17} & 66.74\sd{10.39} & 77.65\sd{6.89} & 64.35\sd{8.26} & 60.95\sd{11.39} & 62.40\sd{9.38} & 69.53\sd{13.02} & 56.99\sd{6.46} & 61.60\sd{5.50} \\

\cmidrule(l){2-11}

& Claude~Code
& \bestcell{99.27\sd{0.64}} & 74.55\sd{0.79} & 85.15\sd{0.29} & \bestcell{77.40\sd{6.29}} & 75.39\sd{5.11} & \bestcell{76.24\sd{4.10}} & 76.78\sd{7.56} & 59.77\sd{5.97} & 67.20\sd{6.58} \\

& mini-SWE-agent
& 93.11\sd{4.57} & 70.16\sd{3.11} & 80.01\sd{3.60} & 63.41\sd{3.36} & 57.97\sd{2.54} & 60.56\sd{2.77} & 74.42\sd{4.16} & 54.00\sd{7.77} & 62.48\sd{6.40} \\

& OpenCode
& 97.90\sd{0.54} & 76.09\sd{2.11} & \bestcell{85.62\sd{1.54}} & 70.73\sd{1.40} & 68.46\sd{1.51} & 69.56\sd{0.57} & 77.63\sd{2.80} & 45.73\sd{2.45} & 57.48\sd{1.27} \\

& \textit{Agent Avg.}
& 96.76\sd{3.23} & 73.60\sd{3.08} & 83.59\sd{3.11} & 70.51\sd{7.00} & 67.27\sd{8.77} & 68.79\sd{7.87} & 76.28\sd{1.66} & 53.17\sd{7.06} & 62.39\sd{4.86} \\

\midrule

\multirow{11}{*}[-3pt]{\rotatebox{90}{\textbf{Python}}} & GPT-5.5
& 88.14\sd{1.00} & 81.43\sd{2.42} & 84.65\sd{1.66} & 71.19\sd{4.17} & \bestcell{73.04\sd{2.67}} & 72.10\sd{3.23} & 71.04\sd{3.68} & \bestcell{73.86\sd{2.13}} & 72.42\sd{3.92} \\

& Claude
& 86.74\sd{1.00} & 83.07\sd{2.42} & 84.86\sd{1.66} & 75.49\sd{4.68} & 70.64\sd{2.99} & 72.99\sd{3.63} & 78.91\sd{3.13} & 69.05\sd{1.81} & 73.65\sd{3.33} \\

& Qwen3.7-Max
& \bestcell{92.49\sd{1.20}} & \bestcell{84.21\sd{2.88}} & \bestcell{88.15\sd{1.98}} & 69.90\sd{5.64} & 67.92\sd{3.61} & 68.90\sd{4.37} & 80.45\sd{4.87} & 71.81\sd{2.81} & \bestcell{75.89\sd{5.18}} \\

& DeepSeek
& 88.62\sd{1.26} & 83.62\sd{3.02} & 86.05\sd{2.08} & 66.02\sd{5.61} & 62.96\sd{3.59} & 64.45\sd{4.36} & 74.31\sd{3.86} & 71.81\sd{2.23} & 73.04\sd{4.10} \\

& Qwen-480B
& 92.13\sd{1.35} & 83.57\sd{3.24} & 87.64\sd{2.23} & \bestcell{77.91\sd{6.17}} & 69.07\sd{3.95} & \bestcell{73.22\sd{4.79}} & 75.68\sd{4.14} & 65.12\sd{2.39} & 70.00\sd{4.40} \\

& Qwen-30B
& 68.18\sd{0.99} & 78.95\sd{2.38} & 73.17\sd{1.64} & 55.56\sd{4.11} & 52.63\sd{2.63} & 54.05\sd{3.19} & \bestcell{81.82\sd{3.45}} & 60.00\sd{1.99} & 69.23\sd{3.67} \\

& \textit{LLM Avg.}
& 86.05\sd{9.05} & 82.48\sd{1.97} & 84.09\sd{5.53} & 69.35\sd{7.95} & 66.04\sd{7.38} & 67.62\sd{7.43} & 77.04\sd{4.08} & 68.61\sd{5.19} & 72.37\sd{2.45} \\

\cmidrule(l){2-11}

& Claude~Code
& 88.39\sd{1.22} & 79.52\sd{4.50} & 83.69\sd{3.03} & 52.83\sd{3.65} & 46.36\sd{3.19} & 49.38\sd{3.39} & 46.80\sd{1.58} & 46.50\sd{1.53} & 46.62\sd{0.31} \\

& mini-SWE-agent
& 89.92\sd{1.49} & 72.87\sd{1.65} & 80.50\sd{1.36} & 54.96\sd{1.04} & 44.21\sd{0.94} & 49.00\sd{0.77} & 49.97\sd{5.13} & 36.63\sd{2.63} & 42.22\sd{3.11} \\

& OpenCode
& 86.91\sd{1.29} & 78.11\sd{1.27} & 82.26\sd{0.27} & 54.36\sd{3.83} & 46.41\sd{2.63} & 50.06\sd{3.09} & 47.21\sd{0.72} & 42.78\sd{0.68} & 44.88\sd{0.23} \\

& \textit{Agent Avg.}
& 88.41\sd{1.51} & 76.83\sd{3.50} & 82.15\sd{1.60} & 54.05\sd{1.10} & 45.66\sd{1.26} & 49.48\sd{0.54} & 47.99\sd{1.72} & 41.97\sd{4.98} & 44.57\sd{2.22} \\

\bottomrule
\end{tabular}%
}
\end{table}

\begin{table}[!tp]
\renewcommand{\arraystretch}{0.82}
\setlength{\abovecaptionskip}{0.1cm}
\caption{RQ2: Evaluation on stubs and mock assertions at a finer-grained aspect.}
\label{tab:rq2_layer}
\centering
\fontsize{6}{7}\selectfont
\renewcommand{\sd}[1]{{\fontsize{5}{6}\selectfont$\pm$#1}}
\setlength{\tabcolsep}{0.2pt}
\renewcommand{\arraystretch}{0.84}
\setlength{\aboverulesep}{0.5pt}
\setlength{\belowrulesep}{0.5pt}

\begin{tabular*}{\linewidth}{@{\extracolsep{\fill}}l@{\hspace{2pt}}l *{12}{c}@{}}
\toprule
& Stub
& \multicolumn{3}{c}{Method}
& \multicolumn{3}{c}{Argument}
& \multicolumn{3}{c}{Response}
& \multicolumn{3}{c}{Output} \\
\cmidrule(lr){3-5}
\cmidrule(lr){6-8}
\cmidrule(lr){9-11}
\cmidrule(lr){12-14}
& Subject
& P & R & F1
& P & R & F1
& P & R & F1
& P & R & F1 \\
\midrule

\multirow{11}{*}[-1pt]{\rotatebox{90}{\textbf{Java}}} & GPT-5.5
& 85.25\sd{6.97} & 87.91\sd{2.08} & 86.52\sd{4.61}
& 76.06\sd{7.34} & 78.40\sd{3.01} & 77.18\sd{5.25}
& 85.25\sd{6.97} & 87.91\sd{2.08} & 86.52\sd{4.61}
& 70.18\sd{6.82} & 72.34\sd{2.83} & 71.21\sd{4.89} \\

& Claude
& 82.93\sd{4.60} & 88.37\sd{5.49} & 85.56\sd{5.03}
& 75.25\sd{5.23} & 80.18\sd{6.11} & 77.64\sd{5.65}
& 82.93\sd{4.60} & 88.37\sd{5.49} & 85.56\sd{5.03}
& 63.79\sd{2.83} & 67.97\sd{3.47} & 65.81\sd{3.13} \\

& DeepSeek
& \bestcell{99.21\sd{1.12}} & 85.19\sd{4.78} & 91.62\sd{2.29}
& \bestcell{93.12\sd{0.75}} & 80.02\sd{6.03} & 86.03\sd{3.81}
& \bestcell{99.21\sd{1.12}} & 85.19\sd{4.78} & 91.62\sd{2.29}
& 75.40\sd{3.37} & 64.68\sd{1.47} & 69.59\sd{0.58} \\

& Qwen3.7-Max
& 90.92\sd{1.17} & 77.85\sd{2.14} & 83.88\sd{1.74}
& 79.58\sd{5.52} & 68.17\sd{5.73} & 73.43\sd{5.67}
& 90.92\sd{1.17} & 77.85\sd{2.14} & 83.88\sd{1.74}
& 70.92\sd{2.20} & 60.74\sd{2.77} & 65.44\sd{2.54} \\

& Qwen-480B
& 89.00\sd{0.67} & \bestcell{93.77\sd{0.95}} & 91.32\sd{0.80}
& 85.61\sd{0.51} & \bestcell{90.20\sd{0.77}} & \bestcell{87.84\sd{0.63}}
& 88.18\sd{1.83} & \bestcell{92.91\sd{2.17}} & 90.48\sd{1.99}
& 77.12\sd{0.10} & 81.26\sd{0.32} & 79.14\sd{0.20} \\

& Qwen-30B
& 88.75\sd{1.77} & 71.71\sd{0.40} & 79.33\sd{0.95}
& 71.25\sd{5.30} & 57.55\sd{3.46} & 63.67\sd{4.24}
& 83.75\sd{1.77} & 67.67\sd{0.46} & 74.86\sd{0.99}
& 70.00\sd{3.54} & 56.55\sd{2.05} & 62.56\sd{2.67} \\

& \textit{LLM Avg.}
& 89.34\sd{5.62} & 84.13\sd{8.00} & 86.37\sd{4.66} & 80.15\sd{7.98} & 75.75\sd{11.33} & 77.63\sd{8.80} & 88.37\sd{6.08} & 83.32\sd{9.13} & 85.49\sd{5.99} & 71.24\sd{4.69} & 67.26\sd{8.79} & 68.96\sd{5.87} \\

\cmidrule(l){2-14}

& Claude~Code
& 94.56\sd{7.85} & 91.98\sd{3.35} & \bestcell{93.07\sd{3.45}}
& 87.69\sd{6.52} & 85.37\sd{4.29} & 86.34\sd{3.08}
& 93.86\sd{7.23} & 91.33\sd{3.63} & \bestcell{92.40\sd{2.91}}
& \bestcell{83.58\sd{6.72}} & \bestcell{81.35\sd{4.41}} & \bestcell{82.29\sd{3.66}} \\

& mini-SWE-agent
& 86.86\sd{1.38} & 79.44\sd{1.71} & 82.97\sd{0.93}
& 76.14\sd{3.05} & 69.60\sd{1.50} & 72.71\sd{2.00}
& 86.41\sd{1.02} & 79.03\sd{1.72} & 82.54\sd{0.73}
& 70.53\sd{3.15} & 64.49\sd{2.66} & 67.37\sd{2.69} \\

& OpenCode
& 90.37\sd{4.19} & 87.40\sd{0.67} & 88.84\sd{2.36}
& 84.07\sd{4.65} & 81.31\sd{2.21} & 82.65\sd{3.18}
& 90.05\sd{4.59} & 87.09\sd{1.09} & 88.52\sd{2.77}
& 75.19\sd{1.86} & 72.76\sd{1.06} & 73.93\sd{0.43} \\

& \textit{Agent Avg.}
& 90.60\sd{3.86} & 86.27\sd{6.35} & 88.29\sd{5.07} & 82.63\sd{5.91} & 78.76\sd{8.19} & 80.57\sd{7.05} & 90.11\sd{3.73} & 85.82\sd{6.25} & 87.82\sd{4.97} & 76.43\sd{6.61} & 72.87\sd{8.43} & 74.53\sd{7.48} \\

\midrule

\multirow{11}{*}[-1pt]{\rotatebox{90}{\textbf{Python}}} & GPT-5.5
& 86.44\sd{1.39} & \bestcell{88.70\sd{2.00}} & 87.55\sd{1.91}
& 85.21\sd{1.42} & \bestcell{87.42\sd{1.93}} & 86.30\sd{2.33}
& 83.05\sd{1.38} & \bestcell{85.22\sd{1.65}} & 84.12\sd{1.64}
& 71.19\sd{4.17} & \bestcell{73.04\sd{2.67}} & 72.10\sd{3.23} \\

& Claude
& 91.18\sd{1.33} & 85.32\sd{1.91} & \bestcell{88.15\sd{1.82}}
& 90.20\sd{3.05} & 84.40\sd{2.86} & \bestcell{87.20\sd{2.95}}
& 88.24\sd{1.69} & 82.57\sd{2.03} & \bestcell{85.31\sd{2.01}}
& 75.49\sd{4.68} & 70.64\sd{2.99} & 72.99\sd{3.63} \\

& DeepSeek
& 86.41\sd{1.53} & 82.41\sd{2.20} & 84.36\sd{2.10}
& 85.44\sd{1.92} & 81.48\sd{1.83} & 83.41\sd{1.87}
& 81.55\sd{2.19} & 77.78\sd{2.63} & 79.62\sd{2.61}
& 66.02\sd{5.61} & 62.96\sd{3.59} & 64.45\sd{4.36} \\

& Qwen3.7-Max
& 88.35\sd{1.78} & 85.85\sd{2.55} & 87.08\sd{2.44}
& 86.41\sd{2.56} & 83.96\sd{2.48} & 85.17\sd{2.52}
& 83.50\sd{2.19} & 81.13\sd{2.62} & 82.30\sd{2.60}
& 69.90\sd{5.64} & 67.92\sd{3.61} & 68.90\sd{4.37} \\

& Qwen-480B
& \bestcell{91.86\sd{1.87}} & 81.44\sd{2.69} & 86.34\sd{2.57}
& \bestcell{90.70\sd{3.04}} & 80.41\sd{2.70} & 85.25\sd{2.86}
& \bestcell{89.53\sd{2.06}} & 79.38\sd{2.46} & 84.15\sd{2.44}
& \bestcell{77.91\sd{6.17}} & 69.07\sd{3.95} & \bestcell{73.22\sd{4.79}} \\

& Qwen-30B
& 83.33\sd{1.50} & 78.95\sd{2.16} & 81.08\sd{2.06}
& 83.26\sd{1.34} & 79.21\sd{1.27} & 81.18\sd{1.30}
& 77.78\sd{1.81} & 73.68\sd{2.17} & 75.68\sd{2.15}
& 55.56\sd{4.11} & 52.63\sd{2.63} & 54.05\sd{3.19} \\

& \textit{LLM Avg.}
& 87.93\sd{3.22} & 83.78\sd{3.51} & 85.76\sd{2.64} & 86.87\sd{2.96} & 82.81\sd{3.02} & 84.75\sd{2.16} & 83.94\sd{4.34} & 79.96\sd{4.01} & 81.86\sd{3.62} & 69.35\sd{7.95} & 66.04\sd{7.38} & 67.62\sd{7.43} \\

\cmidrule(l){2-14}

& Claude~Code
& 75.55\sd{2.57} & 66.29\sd{2.26} & 70.61\sd{2.37}
& 72.93\sd{2.17} & 64.00\sd{2.28} & 68.17\sd{2.22}
& 72.99\sd{3.69} & 64.04\sd{3.17} & 68.23\sd{3.39}
& 52.83\sd{3.65} & 46.36\sd{3.19} & 49.38\sd{3.39} \\

& mini-SWE-agent
& 74.84\sd{1.50} & 60.24\sd{2.75} & 66.74\sd{2.29}
& 72.93\sd{1.70} & 58.71\sd{2.84} & 65.04\sd{2.42}
& 71.92\sd{1.83} & 57.88\sd{2.53} & 64.13\sd{2.22}
& 54.96\sd{1.04} & 44.21\sd{0.94} & 49.00\sd{0.77} \\

& OpenCode
& 76.33\sd{2.63} & 65.18\sd{1.56} & 70.31\sd{1.81}
& 73.74\sd{3.50} & 62.96\sd{2.11} & 67.92\sd{2.59}
& 73.39\sd{2.43} & 62.67\sd{1.32} & 67.60\sd{1.59}
& 54.36\sd{3.83} & 46.41\sd{2.63} & 50.06\sd{3.09} \\

& \textit{Agent Avg.}
& 75.57\sd{0.75} & 63.90\sd{3.22} & 69.22\sd{2.15} & 73.20\sd{0.47} & 61.89\sd{2.80} & 67.04\sd{1.74} & 72.77\sd{0.76} & 61.53\sd{3.23} & 66.65\sd{2.21} & 54.05\sd{1.10} & 45.66\sd{1.26} & 49.48\sd{0.54} \\

\bottomrule
\end{tabular*}

\begin{tabular*}{\linewidth}{@{\extracolsep{\fill}}l@{\hspace{2pt}}l *{9}{c}@{}}
\toprule
& Mock assertion
& \multicolumn{3}{c}{Method}
& \multicolumn{3}{c}{Argument}
& \multicolumn{3}{c}{Constraint} \\
\cmidrule(lr){3-5}
\cmidrule(lr){6-8}
\cmidrule(lr){9-11}
& Subject
& P & R & F1
& P & R & F1
& P & R & F1 \\
\midrule

\multirow{11}{*}[-1pt]{\rotatebox{90}{\textbf{Java}}} & GPT-5.5
& 56.26\sd{3.42} & \bestcell{77.49\sd{1.98}} & 65.18\sd{2.99}
& 49.08\sd{3.05} & \bestcell{67.60\sd{1.83}} & 56.86\sd{2.70}
& 54.28\sd{3.88} & \bestcell{74.75\sd{2.72}} & 62.88\sd{3.57} \\

& Claude
& 77.20\sd{10.33} & 72.23\sd{1.14} & 74.42\sd{4.24}
& 70.99\sd{9.16} & 66.45\sd{1.37} & 68.45\sd{3.56}
& 69.38\sd{11.44} & 64.76\sd{1.01} & 66.80\sd{5.89} \\

& DeepSeek
& 94.44\sd{7.86} & 69.18\sd{9.92} & \bestcell{79.36\sd{3.81}}
& 84.22\sd{3.87} & 61.95\sd{11.16} & 70.94\sd{6.04}
& 90.42\sd{7.40} & 66.24\sd{9.59} & 75.98\sd{3.75} \\

& Qwen3.7-Max
& 87.42\sd{0.65} & 65.98\sd{2.07} & 75.19\sd{1.59}
& 73.29\sd{2.93} & 55.28\sd{0.89} & 63.02\sd{1.67}
& 82.21\sd{0.56} & 62.04\sd{1.06} & 70.71\sd{0.48} \\

& Qwen-480B
& 93.45\sd{1.62} & 63.62\sd{6.08} & 75.59\sd{3.78}
& 85.55\sd{1.32} & 58.25\sd{5.68} & 69.20\sd{3.59}
& 86.83\sd{0.49} & 59.19\sd{7.01} & 70.29\sd{5.12} \\

& Qwen-30B
& \bestcell{96.67\sd{0.00}} & 64.44\sd{0.00} & 77.33\sd{0.00}
& 71.67\sd{2.36} & 47.78\sd{1.57} & 57.33\sd{1.89}
& \bestcell{95.00\sd{2.36}} & 63.33\sd{1.57} & \bestcell{76.00\sd{1.89}} \\

& \textit{LLM Avg.}
& 84.24\sd{15.40} & 68.82\sd{5.31} & 74.51\sd{4.90} & 72.47\sd{13.11} & 59.55\sd{7.44} & 64.30\sd{6.18} & 79.69\sd{15.22} & 65.05\sd{5.33} & 70.44\sd{5.14} \\

\cmidrule(l){2-11}

& Claude~Code
& 88.96\sd{4.19} & 69.23\sd{2.59} & 77.85\sd{3.03}
& 82.32\sd{5.22} & 64.07\sd{3.93} & \bestcell{72.05\sd{4.33}}
& 83.42\sd{6.17} & 64.92\sd{4.63} & 73.01\sd{5.15} \\

& mini-SWE-agent
& 87.46\sd{3.93} & 63.20\sd{5.41} & 73.25\sd{3.66}
& 78.33\sd{3.29} & 56.76\sd{7.13} & 65.71\sd{5.54}
& 82.25\sd{3.69} & 59.49\sd{5.94} & 68.92\sd{4.35} \\

& OpenCode
& 92.73\sd{2.63} & 54.64\sd{3.30} & 68.68\sd{1.96}
& \bestcell{85.81\sd{3.15}} & 50.54\sd{2.61} & 63.54\sd{1.25}
& 83.83\sd{1.11} & 49.45\sd{3.75} & 62.13\sd{2.74} \\

& \textit{Agent Avg.}
& 89.72\sd{2.72} & 62.36\sd{7.33} & 73.26\sd{4.59} & 82.15\sd{3.74} & 57.12\sd{6.77} & 67.10\sd{4.42} & 83.17\sd{0.82} & 57.95\sd{7.85} & 68.02\sd{5.50} \\

\midrule

\multirow{11}{*}[-1pt]{\rotatebox{90}{\textbf{Python}}} & GPT-5.5
& 86.89\sd{2.01} & \bestcell{90.34\sd{1.62}} & 88.58\sd{2.70}
& 75.96\sd{2.30} & 78.98\sd{1.31} & 77.44\sd{2.40}
& 74.32\sd{2.47} & \bestcell{77.27\sd{1.70}} & 75.77\sd{2.90} \\

& Claude
& 93.88\sd{2.15} & 82.14\sd{1.73} & 87.62\sd{2.89}
& 84.35\sd{2.54} & 73.81\sd{1.44} & 78.73\sd{2.65}
& \bestcell{85.03\sd{3.15}} & 74.40\sd{2.17} & \bestcell{79.37\sd{3.70}} \\

& DeepSeek
& 91.67\sd{3.08} & 88.59\sd{2.48} & \bestcell{90.10\sd{4.15}}
& 80.56\sd{3.13} & 77.85\sd{1.77} & 79.18\sd{3.27}
& 78.47\sd{3.49} & 75.84\sd{2.40} & 77.13\sd{4.10} \\

& Qwen3.7-Max
& \bestcell{95.49\sd{2.41}} & 85.23\sd{1.94} & 90.07\sd{3.25}
& \bestcell{89.47\sd{2.97}} & \bestcell{79.87\sd{1.69}} & \bestcell{84.40\sd{3.11}}
& 82.71\sd{3.13} & 73.83\sd{2.15} & 78.01\sd{3.67} \\

& Qwen-480B
& 92.79\sd{2.59} & 79.84\sd{2.08} & 85.83\sd{3.49}
& 84.68\sd{3.20} & 72.87\sd{1.82} & 78.33\sd{3.34}
& 78.38\sd{3.98} & 67.44\sd{2.73} & 72.50\sd{4.66} \\

& Qwen-30B
& 90.91\sd{2.09} & 66.67\sd{1.68} & 76.92\sd{2.82}
& 81.82\sd{2.56} & 60.00\sd{1.45} & 69.23\sd{2.67}
& 81.82\sd{2.45} & 60.00\sd{1.68} & 69.23\sd{2.87} \\

& \textit{LLM Avg.}
& 91.94\sd{2.96} & 82.14\sd{8.52} & 86.52\sd{4.97} & 82.81\sd{4.54} & 73.90\sd{7.37} & 77.89\sd{4.90} & 80.12\sd{3.82} & 71.46\sd{6.55} & 75.34\sd{3.80} \\

\cmidrule(l){2-11}

& Claude~Code
& 67.56\sd{1.53} & 67.20\sd{4.15} & 67.33\sd{2.15}
& 56.90\sd{1.10} & 56.57\sd{2.81} & 56.69\sd{1.03}
& 61.48\sd{1.12} & 61.12\sd{3.03} & 61.25\sd{1.06} \\

& mini-SWE-agent
& 70.97\sd{5.51} & 52.07\sd{2.96} & 59.99\sd{2.98}
& 57.76\sd{4.29} & 42.38\sd{2.07} & 48.83\sd{2.05}
& 63.54\sd{4.62} & 46.61\sd{1.99} & 53.71\sd{1.97} \\

& OpenCode
& 66.17\sd{3.87} & 59.91\sd{1.90} & 62.87\sd{2.78}
& 55.65\sd{3.72} & 50.37\sd{1.92} & 52.87\sd{2.74}
& 59.94\sd{4.37} & 54.26\sd{2.51} & 56.95\sd{3.35} \\

& \textit{Agent Avg.}
& 68.23\sd{2.47} & 59.73\sd{7.57} & 63.40\sd{3.70} & 56.77\sd{1.06} & 49.77\sd{7.11} & 52.80\sd{3.93} & 61.65\sd{1.81} & 54.00\sd{7.26} & 57.30\sd{3.78} \\

\bottomrule
\end{tabular*}
\end{table}

\textbf{Comparison among mock artifacts.} 
Across both languages, mock objects consistently outperform stubs and mock assertions across all models and agents at the coarse-grained level (Table~\ref{tab:rq2_component}), leading by 6.58--25.24 pp in Precision, 4.28--28.39 pp in Recall, and 6.10--23.12 pp in F1.
Because mock object declarations only need to identify the proper dependencies to replace,
whereas stubs and mock assertions must additionally specify how these dependencies respond and which interactions should be checked.

To dive deeper into a fine-grained evaluation of stub and mock assertion quality, we present Table~\ref{tab:rq2_layer} with assessment in terms of Method, Argument, Response, and Output for stubs, as well as Method, Argument, and Constraint for mock assertions. 
For stubs, the correctness of Output consistently achieves the lowest precision, recall, and F1-score among the four evaluation aspects.
Across standalone LLMs and coding agents, average Output F1 trails Method, Response, and Argument F1 by 6.04--17.41 pp on Java and 14.24--19.74 pp on Python, making Output the weakest stub aspect in both languages.
These results indicate that the major difficulty in stub generation lies in determining the concrete output of a stub. 
While methods and arguments can often be directly extracted from dependency calls in the MUT and response types can often be inferred from surrounding control structures (e.g., \texttt{try-catch} blocks or assignment statements), concrete outputs do not appear in the MUT. Instead, they must be tailored to the specific interaction logic intended by the test scenario, which requires models to understand how the returned state drives the execution path.
Moreover, although coding agents can iteratively repair tests via runtime feedback, execution errors (e.g., generic \texttt{NullPointerException} or assertion failures) rarely convey the exact expected return values or exceptions, leaving Output synthesis a fundamental challenge across both
standalone LLMs and coding agents.
  


For mock assertions, invocation methods consistently achieve higher F1-score than arguments and invocation-count constraints across both standalone LLMs and coding agents(Table~\ref{tab:rq2_layer}).
On average, Argument F1-score falls behind Method F1-score by 6.16--10.21 pp on Java and 8.64--10.60 pp on Python across LLMs and coding agents, while Constraint F1-score lags by 4.07--5.24 pp on Java and 6.10--11.19 pp on Python.
These discrepancies indicate that identifying which method to verify is much easier than verifying how it is called.
Specifically, identifying the invocation target mainly requires locating which collaborator method is called, whereas reproducing its arguments additionally requires tracing how values are computed and transformed inside the MUT. These arguments may be assembled from multiple variables, converted into new representations, or serialized before the collaborator is invoked.
When models are uncertain about these intermediate values, they often resort to permissive matchers like \texttt{any()}, leading to low argument fidelity.
Likewise, verifying the exact invocation count requires accurately tracking loops and conditional branches, which models frequently misjudge even when they correctly identify the target method.

\begin{findingbox}
\textbf{Finding 4:}
Compared with mock objects, LLMs and coding agents have more difficulty in stub and mock assertion generation.  Dive deeper into the latter two, we found that Output of stubs and Argument, alongside Constraint, of mock assertions are much harder for correct generation, owing to a comprehensive understanding of the code.
\end{findingbox}

\textbf{Does feedback repair of coding agents improve mock quality?}
While coding agents leverage iterative test execution feedback to resolve runtime failures, Table~\ref{tab:rq2_component} reveals that this feedback mechanism does not translate into higher mock fidelity over standalone LLMs.
Specifically, standalone LLMs outperform coding agents by an average of 18.14 pp and 27.80 pp in F1 score for stubs and mock assertions on Python, while on Java, agents yield only a marginal 6.38 pp gain on stubs and show no advantage on mock assertions.
A Zipkin task illustrates how repair can make a test pass without restoring the intended interaction constraint. The reference test requires exactly two invocations of \texttt{getCassandraVersion()}. Claude Code initially generated \texttt{verify(node).getCassandraVersion()}, which checks for one invocation and caused the test to fail. The agent then changed the assertion to \texttt{atLeastOnce()} rather than restoring the exact count. The repaired test passed, but the assertion could no longer detect calls beyond the required two.
This case illustrates the distinction between executability and behavior fidelity. Compiler and test feedback help agents identify errors, but passing the available tests does not establish that generated mocks preserve all intended collaborator behaviors. Consequently, successful repair alone is insufficient to demonstrate improved behavior fidelity.

\begin{findingbox}
\textbf{Finding 5:}
Although iterative test execution feedback enables coding agents to improve test executability, it fails to enhance mock behavior fidelity, yielding no overall advantage in mock quality over standalone LLMs.
\end{findingbox}
\subsection{RQ3: Interaction-Fault Detection Capability}\label{sec:rq3}

\textbf{Fault detection under mutation analysis.} 
Table~\ref{tab:rq3mutation} summarizes the results across both languages.
Overall, no LLM or coding agent fully preserves the fault-detection capability of $T^G$, with average Jaccard similarities ranging from 70.85\% on Java to 85.87\% on Python.
Consistent with our observations in RQ2~\ref{sec:rq2}, all evaluated LLMs and coding agents achieve significantly higher Jaccard similarity on Python than on Java, which is driven by the superior fidelity of stub and mock assertion generation in Python (e.g., GPT-5.5 reaches 95.83\% on Python vs. 90.82\% on Java).

More importantly, coding agents exhibit a noticeable drop in mutation kill similarity compared to top-tier standalone LLMs.
On Java, while standalone LLMs like GPT-5.5 achieve up to 90.82\% similarity, coding agents only attain 60.04\%--70.74\% (e.g., OpenCode drops to 60.04\% and mini-SWE-agent to 64.22\%). 
This phenomenon directly aligns with our findings in Section~\ref{sec:rq2}: as revealed in Table~\ref{tab:rq2_layer} and the repair analysis, coding agents may game runtime feedback by relaxing or omitting mock assertions (e.g., weakening call counts) to pass the test runner. 


\begingroup
\setlength{\intextsep}{4pt}
\setlength{\textfloatsep}{4pt}
\renewcommand{\arraystretch}{0.84}
\setlength{\aboverulesep}{0.5pt}
\setlength{\belowrulesep}{0.5pt}
\begin{table}[H]
\captionsetup{skip=3pt}
\caption{Kill-set Jaccard similarity $J$ (\%) of passing $T^C$.}
\label{tab:rq3mutation}
\centering
\scriptsize
\setlength{\tabcolsep}{2pt}
\renewcommand{\arraystretch}{1.08}
\begin{tabular*}{\linewidth}{@{\extracolsep{\fill}}l*{9}{c}@{}}
\toprule
& \multicolumn{6}{c}{\textit{Standalone LLMs}}
& \multicolumn{3}{c}{\textit{Coding agents}} \\
\cmidrule(lr){2-7}\cmidrule(l){8-10}
\textbf{Language}
& \makecell{GPT-5.5}
& \makecell{Claude}
& \makecell{Qwen3.7-\\Max}
& \makecell{DeepSeek    }
& \makecell{Qwen-480B}
& \makecell{Qwen-30B}
& \makecell{Claude\\Code}
& \makecell{mini-SWE-\\agent}
& \makecell{OpenCode} \\
\midrule
Java
& 90.82\sd{2.73} & 75.04\sd{5.42} & 72.37\sd{2.27}
& 68.85\sd{10.01} & 68.89\sd{3.14} & 66.67\sd{0.00}
& 70.74\sd{2.91} & 64.22\sd{10.14} & 60.04\sd{7.39} \\
Python
& 95.83\sd{1.85} & 88.78\sd{1.42} & 89.74\sd{2.54}
& 97.50\sd{2.20} & 79.37\sd{1.40} & 80.00\sd{1.09}
& 89.82\sd{2.32} & 69.88\sd{3.92} & 81.88\sd{2.32} \\
\bottomrule
\end{tabular*}
\end{table}
\endgroup

\begin{table}[b]
\centering
\setlength{\abovecaptionskip}{2pt}
\setlength{\belowcaptionskip}{2pt}
\caption{MyBatis case: interaction verification gap between $T^G$ and $T^C$.}
\label{tab:rq3-mybatis-case}

\fontsize{7.5}{8.2}\selectfont
\setlength{\tabcolsep}{5pt}
\renewcommand{\arraystretch}{0.90}
\setlength{\aboverulesep}{1pt}
\setlength{\belowrulesep}{1pt}

\begin{tabular*}{\linewidth}{@{\extracolsep{\fill}}llcc@{}}
\toprule
\textbf{Suite} &
\multicolumn{3}{l@{}}{\textbf{Verification Logic (Mock Assertions)}} \\
$T^G$ &
\multicolumn{3}{l@{}}{\texttt{verify(cs).getObject(idx, ...); verify(cs, \textbf{never()}).wasNull();}} \\
$T^C$ &
\multicolumn{3}{l@{}}{
\texttt{verify(cs).getObject(idx, ...);}
\quad \textcolor{red}{\textbf{(no \texttt{wasNull()} check)}}
} \\
\midrule
\textbf{Program Version} &
\textbf{Interaction Behavior / Injected Fault} &
\textbf{$T^G$} &
\textbf{$T^C$} \\
\midrule
Original &
Normal: \texttt{return cs.getObject(idx, OffsetDateTime.class);} &
Pass & Pass \\
Mutant 1 &
\texttt{cs.getObject(idx, ...); return cs.getObject(idx, ...);} &
Killed & Killed \\
Mutant 2 &
\texttt{cs.wasNull(); return cs.getObject(idx, ...);} &
Killed & \textcolor{red}{\textbf{Live}} \\
\bottomrule
\end{tabular*}
\end{table}
\textbf{Case study on mutation-based interaction verification.}
The MyBatis case (Table~\ref{tab:rq3-mybatis-case}) illustrates this verification gap.
$T^G$ verifies two interactions: invoking \texttt{getObject} and strictly ensuring \texttt{wasNull} is never called.
In contrast, $T^C$ only checks \texttt{getObject}, omitting the \texttt{never()} constraint.
During normal execution, both $T^G$ and $T^C$ pass, completely masking this deficiency.
However, mutation testing exposes the gap: while both kill Mutant~1 (a duplicated \texttt{getObject} call), only $T^G$ catches Mutant~2 because $T^C$ fails to detect the forbidden \texttt{wasNull} call.

\begin{findingbox}
\textbf{Finding 6:}
In mutation testing, coding agents exhibit no advantage over standalone LLMs in defect detection capability, as iterative repair often trades verification strictness for test execution, masking subtle interaction faults caught by reference tests.
\end{findingbox}

As shown in Table~\ref{tab:rq4}, \textsc{StubCoder} and the LLM-based approaches exhibit distinct strengths and failure modes:
First, regarding executability, all three coding agents achieve 100\% TPR, consistently outperforming \textsc{StubCoder}, which achieves 74.6\% when reproduced under its original generation-mode configuration (Table~\ref{tab:rq4}).
However, standalone DeepSeek achieves only 61.0\%, lagging behind \textsc{StubCoder}.
This highlights that without runtime feedback, standalone LLMs struggle to ensure the syntactical and type validity of stub codes, falling behind type-guided, search-based tools like \textsc{StubCoder} that operate over statically resolved symbol spaces.
Coding agent bridge this gap. Instead of guessing in a single pass, coding agents read the compiler errors and test failures, step-by-step correcting import and syntax mistakes until all suites execute successfully.

As for the stub quality, running without errors still does not guarantee that the generated stubs are correct.
Following the protocol in Section~\ref{sec:evaluation_metrics}, we evaluate the fine-grained fidelity of the synthesized stub code across four dimensions (Table~\ref{tab:rq4}).
Across these dimensions, both standalone DeepSeek and the three agents consistently outperform \textsc{StubCoder}.
The widest gap appears in argument reconstruction: while standalone DeepSeek and the coding agents achieve F1 scores between 0.615 and 0.787, \textsc{StubCoder} manages only 0.126.
This disparity stems from their underlying synthesis mechanisms. 
\textsc{StubCoder} relies on search-based techniques over a symbol pool mined strictly from local test contexts. 
While this localized pool suffices for identifying visible mock targets and method signatures, it cannot capture the interaction logic between the caller and its dependencies, failing to synthesize complex object states required by subsequent operations. 
In contrast, LLM-based approaches leverage vast BPE-based vocabularies to flexibly synthesize out-of-context objects and variables beyond a fixed local symbol pool.

\begin{findingbox}
\textbf{Finding 7:} 
Overall, LLM-based approaches, regardless of standalone LLM or coding agents, have outperformed traditional stub generation tools, especially in reproducing stub behavior fidelity.
\end{findingbox}
\begingroup
\setlength{\intextsep}{2pt plus 0.5pt minus 0.5pt}

\begin{table}[H]
\centering
\captionsetup{position=top,skip=1pt}
\caption{RQ4: Comparison with StubCoder}
\label{tab:rq4}

\footnotesize
\setlength{\tabcolsep}{3.5pt}
\renewcommand{\arraystretch}{0.88}
\setlength{\aboverulesep}{0.6pt}
\setlength{\belowrulesep}{0.6pt}

\begin{tabular*}{\linewidth}{@{\extracolsep{\fill}}lccccc@{}}
\toprule
Approach & TPR (\%) & Method F1 & Argument F1 & Response F1 & Output F1 \\
\midrule
\textsc{StubCoder}~\cite{zhu2023stubcoder}
& 74.6 & 0.402 & 0.126 & 0.394 & 0.165 \\
DeepSeek
& 61.0 & 0.759 & 0.615 & 0.759 & 0.483 \\
\addlinespace[0.5pt]
Claude~Code
& \bestcell{100.0} & 0.763 & 0.669 & 0.732 & 0.498 \\
mini-SWE-agent
& \bestcell{100.0} & 0.803 & 0.689 & 0.798 & 0.546 \\
OpenCode
& \bestcell{100.0} & \bestcell{0.837} & \bestcell{0.787} & \bestcell{0.828} & \bestcell{0.644} \\
\bottomrule
\multicolumn{6}{@{}l}{\scriptsize All agents use the DeepSeek backbone.}
\end{tabular*}
\end{table}

\endgroup
\subsection{RQ5: Resource Sensitivity}\label{sec:rq5}
\begin{figure*}[t]
  \centering
  \includegraphics[width=0.92\textwidth]{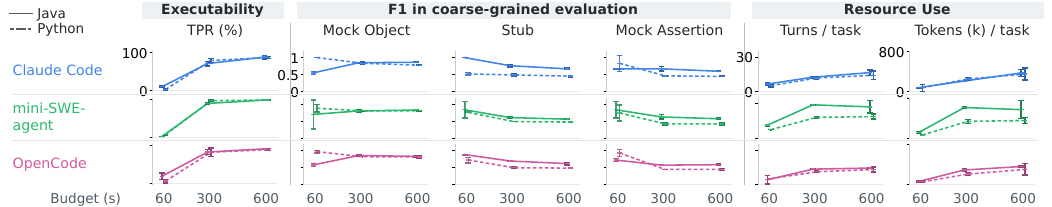}
  \caption{Performance across time budgets.}
  \label{fig:rq5_sensitivity}
  \Description{Comparison of coding-agent executability and mock behavior fidelity across 60, 300, and 600 second time budgets for Java and Python.}
\end{figure*}                         
\noindent\textbf{Executability under different time budgets.}
Increasing the wall-clock budget substantially improves the executability of $T^C$ across all three coding agents and both languages(Figure~\ref{fig:rq5_sensitivity}).
At 60\,s, the average TPR across the three agents is only 10.5\% on Java and 5.5\% on Python.
When the budget is extended to 600\,s, these values increase to 92.4\% and 91.4\%, respectively.
The sharp improvement indicates that a short time budget frequently prevents agents from completing the generation--validation--repair workflow.
In other words, a considerable fraction of execution failures under tight budgets stems from insufficient time to complete the task rather than from an inability to produce executable tests.

\textbf{Resource cost of additional execution time.}
The higher executability is accompanied by greater agent activity and token consumption (Figure~\ref{fig:rq5_sensitivity}).
For example, the average number of interaction rounds across the three agents increases from 7.5 at 60\,s to 18.4 at 300\,s on Java, and from 6.1 to 13.8 on Python.
However, resource consumption does not grow proportionally with the wall-clock limit.
From 300\,s to 600\,s, the increase in interaction rounds is much smaller, reaching 19.5 rounds on Java and 15.5 on Python.
A similar pattern appears in token usage: some agents consume only moderately more tokens under 600\,s, while mini-SWE-agent shows almost no additional consumption on Python and even a lower mean on Java.
This is expected because the wall-clock budget is an upper bound: agents may terminate before exhausting the available time, and different execution trajectories can consume different amounts of computation.
\looseness=-1
\textbf{Behavioral fidelity under larger budgets.}
The additional computation, however, does not yield corresponding improvements in mock behavioral fidelity (Figure~\ref{fig:rq5_sensitivity}).
At the main 300\,s setting, the average agent coarse-grained F1 scores for mock objects, stubs, and mock assertions are 83.59\%, 68.78\%, and 62.39\% on Java, respectively.
At 600\,s, these scores remain at 84.02\%, 61.68\%, and 58.73\%.
The same pattern appears on Python: the corresponding F1 scores change from 82.15\%, 49.48\%, and 44.57\% at 300\,s to 79.84\%, 47.30\%, and 44.05\% at 600\,s.
Thus, although a larger budget allows substantially more candidates to execute successfully, the fidelity gaps identified in RQ2, particularly for stubs and mock assertions, persist.
We do not interpret the comparatively high F1 scores at 60\,s as evidence that shorter budgets produce better mocks, since only a very small, selective subset of candidates becomes executable within 60\,s for fidelity evaluation.

\begin{findingbox}
\textbf{Finding 8:}
The performance of coding agents is sensitive to the available wall-clock budget.
Larger budgets substantially improve executability at increased resource cost, but these gains do not translate into corresponding improvements in mock behavioral fidelity.
\end{findingbox}

\section{Discussion}\label{sec:discussion}

\subsection{Failure Analysis}\label{sec:failure-analysis}
This section invested approximately 120 person-hours in a systematic qualitative study following the thematic analysis guidelines by Cruzes and Dyb{\aa}~\cite{cruzes2011recommended}.
We analyzed 660 executable $T^{C}$s from a single evaluation run across standalone DeepSeek and three DeepSeek-powered coding agents. 
Recall the experimental results in RQ2: no single LLM or coding agent consistently surpasses others. Thus, we chose DeepSeek and used it as a backbone across agents in this RQ for its cost-effectiveness.

First, a domain expert with over six years of development experience in Python and Java analyzes a random 20\% sample of the failure cases to identify and summarize the underlying causes.
The expert iteratively grouped these observations into well-defined error categories, establishing clear taxonomy criteria and representative examples, which were formalized into a codebook.
Next, we recruited four independent evaluators, all with computer science backgrounds and at least three years of professional or research experience in Python and Java.
The evaluation followed a rigorous, double-blind, dual-review process: each sample was inspected independently by two evaluators without cross-communication.
When both evaluators agreed on the classification, the result was finalized (Cohen's Kappa $\kappa = 0.794$).
Disagreements were resolved through structured discussion; if a consensus could not be reached, the case was escalated to the domain expert for final arbitration.

\begin{table}[!htbp]
\centering
\setlength{\abovecaptionskip}{0.1cm}
\caption{Types and frequencies of mock-artifact reconstruction failures.}
\label{tab:reference-loss-types}

\scriptsize
\setlength{\tabcolsep}{2pt}
\renewcommand{\arraystretch}{0.9}
\setlength{\aboverulesep}{0pt}
\setlength{\belowrulesep}{0pt}

\def\failurecolwidth{\dimexpr(\linewidth-28\tabcolsep)/15\relax}

\begin{threeparttable}
\begin{tabular}{@{}
  *{2}{>{\raggedright\arraybackslash}p{\failurecolwidth}}
  >{\centering\arraybackslash
    \columncolor{blue!22}}p{\failurecolwidth}
  *{2}{>{\centering\arraybackslash
         \columncolor{blue!9}}p{\failurecolwidth}}
  >{\centering\arraybackslash
    \columncolor{green!22}}p{\failurecolwidth}
  *{3}{>{\centering\arraybackslash
         \columncolor{green!9}}p{\failurecolwidth}}
  >{\centering\arraybackslash
    \columncolor{purple!22}}p{\failurecolwidth}
  *{5}{>{\centering\arraybackslash
         \columncolor{purple!9}}p{\failurecolwidth}}
  @{}}
\toprule
\multicolumn{2}{@{}l}{Failure type (\%)}
& \textbf{A} & A-1 & A-2
& \textbf{B} & B-1 & B-2 & B-3
& \textbf{C} & C-1 & C-2 & C-3 & C-4 & C-5 \\
\midrule
\multirow{2}{*}{\textbf{Java}} & LLM
& 14.29 & 0.00 & 14.29
& \textbf{71.43} & 42.86 & 14.29 & 28.57
& 42.86 & 14.29 & 28.57 & 0.00 & 0.00 & 0.00 \\
& Agent
& 37.04 & 11.11 & 25.93
& 59.26 & 29.63 & 22.22 & 18.52
& \textbf{66.67} & 44.44 & 22.22 & 11.11 & 0.00 & 0.00 \\
\midrule
\multirow{2}{*}{\textbf{Python}} & LLM
& 18.75 & 0.00 & 18.75
& \textbf{68.75} & 25.00 & 0.00 & 43.75
& 50.00 & 18.75 & 43.75 & 6.25 & 0.00 & 0.00 \\
& Agent
& 29.23 & 10.77 & 20.00
& 55.38 & 15.38 & 3.08 & 40.00
& \textbf{81.54} & 67.69 & 13.85 & 4.62 & 4.62 & 1.54 \\
\bottomrule
\end{tabular}
\end{threeparttable}
\end{table}


Table~\ref{tab:reference-loss-types} summarizes the failure types across executable candidate suites, revealing distinct error distributions between standalone LLMs and coding agents:
\textbf{(1) Standalone LLMs primarily suffer from stubbing errors} (71.43\% in Java and 68.75\% in Python), as direct synthesis lacking execution feedback often yields incorrect stub behaviors or response values.
\textbf{(2) Coding agents fail predominantly from mock assertion errors} (66.67\% in Java and 81.54\% in Python), with missing assertions alone exceeding 44\% across languages. This explains our finding in RQ2 (Section~\ref{sec:rq2}): feedback-driven repair primarily optimizes for passing test execution, where agents often eliminate runtime failures by weakening or discarding strict interaction checks rather than fixing them.
\textbf{(3) Mock object errors are the least frequent} (14.29\%--37.04\%), showing that identifying target dependencies is relatively straightforward compared to generating correct logic and verifications. 
We illustrate each type of error below.

\begin{figure}[!htbp]
\centering
\setlength{\abovecaptionskip}{2pt}
\includegraphics[width=0.95\linewidth]
{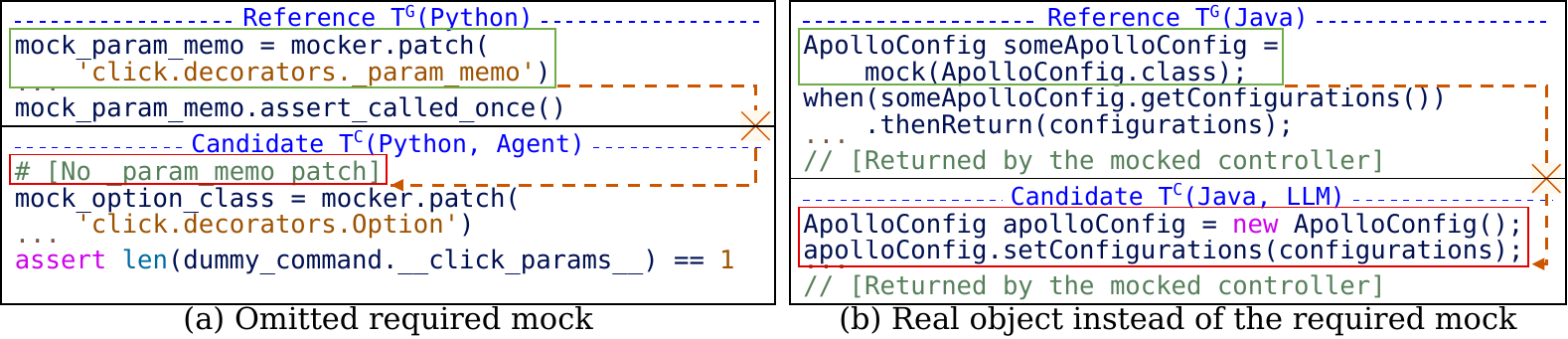}
\caption{Examples of Required mock objects not reconstructed.}
\label{fig:reference-loss-objects}
\Description{Reference and candidate code comparisons showing an
omitted Click helper mock and an ApolloConfig mock replaced with a
real ApolloConfig instance.}
\end{figure}

\textbf{Mock object errors(A).}
\textbf{(1) Missing mocks(A-1):}
This error occurs when $T^C$ completely omits a dependency mocked in $T^G$, breaking test isolation.
As Figure~\ref{fig:reference-loss-objects}(a) shows, $T^C$ mocks \texttt{Option} while omitting the dependency \texttt{\_param\_memo} mocked in $T^G$.
As a result, the real \texttt{\_param\_memo} remains active during execution. This error turns the intended isolated unit test into an accidental integration test, making it fragile to underlying implementation changes. 
\textbf{(2) Non-mock replacements(A-2):}
Here, models avoid mocking altogether by creating real, concrete objects instead of dynamic mocks.
Figure~\ref{fig:reference-loss-objects}(b) shows this in the Apollo case: instead of mocking \texttt{ApolloConfig} to intercept \texttt{getConfigurations()}, the generator creates a real configuration object and calls \texttt{setConfigurations(configurations)}.
This real object provides enough in-memory state to prevent runtime crashes, but it strips away the interaction semantics required by $T^G$.

\begin{figure}[!htbp]
\centering
\setlength{\abovecaptionskip}{2pt}
\includegraphics[width=0.95\linewidth]
{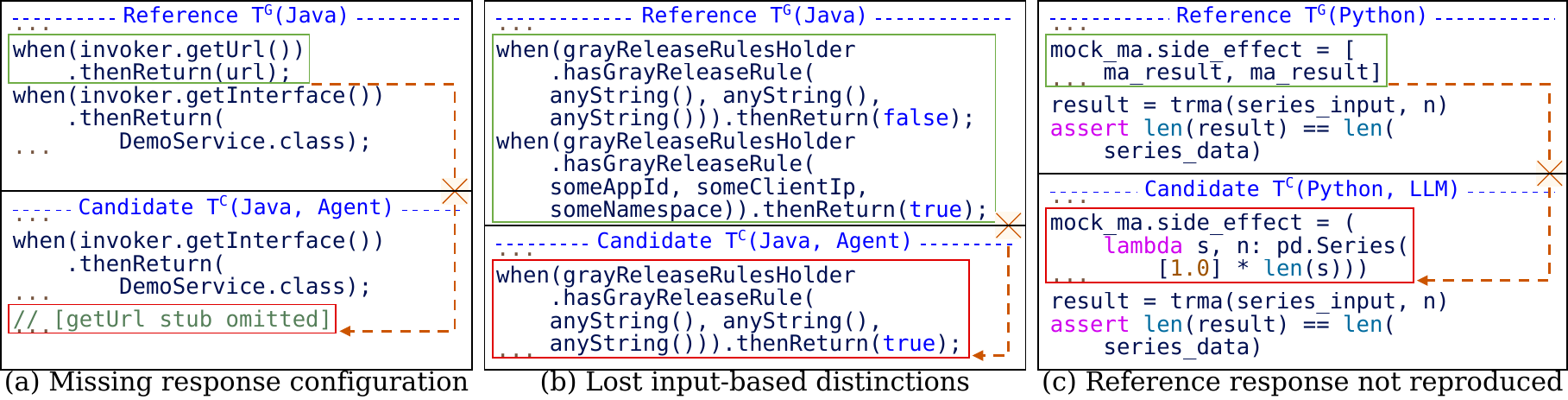}
\caption{Examples of unretained stub specifications.}
\label{fig:reference-loss-stubs}
\Description{Three reference and candidate code comparisons showing
a missing getUrl response in Dubbo, input-dependent gray-release
responses collapsed to true in Apollo, and a reference series
replaced with an all-ones response in Tqsdk.
Green boxes identify reference specifications, and red boxes identify
their missing or altered counterparts.}
\end{figure}

\textbf{Stubbing errors(B).}
\textbf{(1) Missing stubs(B-1):}
A mock is retained, but one of its reference-specified stubs is omitted.
As shown in Figure~\ref{fig:reference-loss-stubs}(a), while the reference test configures responses for \texttt{getInterface}, \texttt{invoke}, and \texttt{invoker.getUrl()}, the agent preserves the former two but drops \texttt{getUrl()}.
\textbf{(2)Argument-matching errors(B-2):}
Multiple responses conditioned on distinct input parameters collapse into a single catch-all stub.
In Apollo's reference gray-release test, the stub returns \texttt{false} for arbitrary string inputs but \texttt{true} for a specific argument combination.
The agent collapses this conditional logic into a broadly matched \texttt{true} response (Figure~\ref{fig:reference-loss-stubs}(b)).
\textbf{(3) Stub behavior errors(B-3):}
A response is configured for the intended call, but its content diverges from the reference values.
In Tqsdk, the reference test configures the mocked \texttt{ma} to return a scenario-derived series (\texttt{ma\_result}) containing specific numeric values and NaNs across two invocations.
Instead of reproducing these realistic values, the agent supplies a trivial all-ones series via \texttt{side\_effect} (Figure~\ref{fig:reference-loss-stubs}(c)).

\begin{figure}[!htbp]
\centering
\setlength{\abovecaptionskip}{2pt}
\includegraphics[width=0.95\linewidth]
{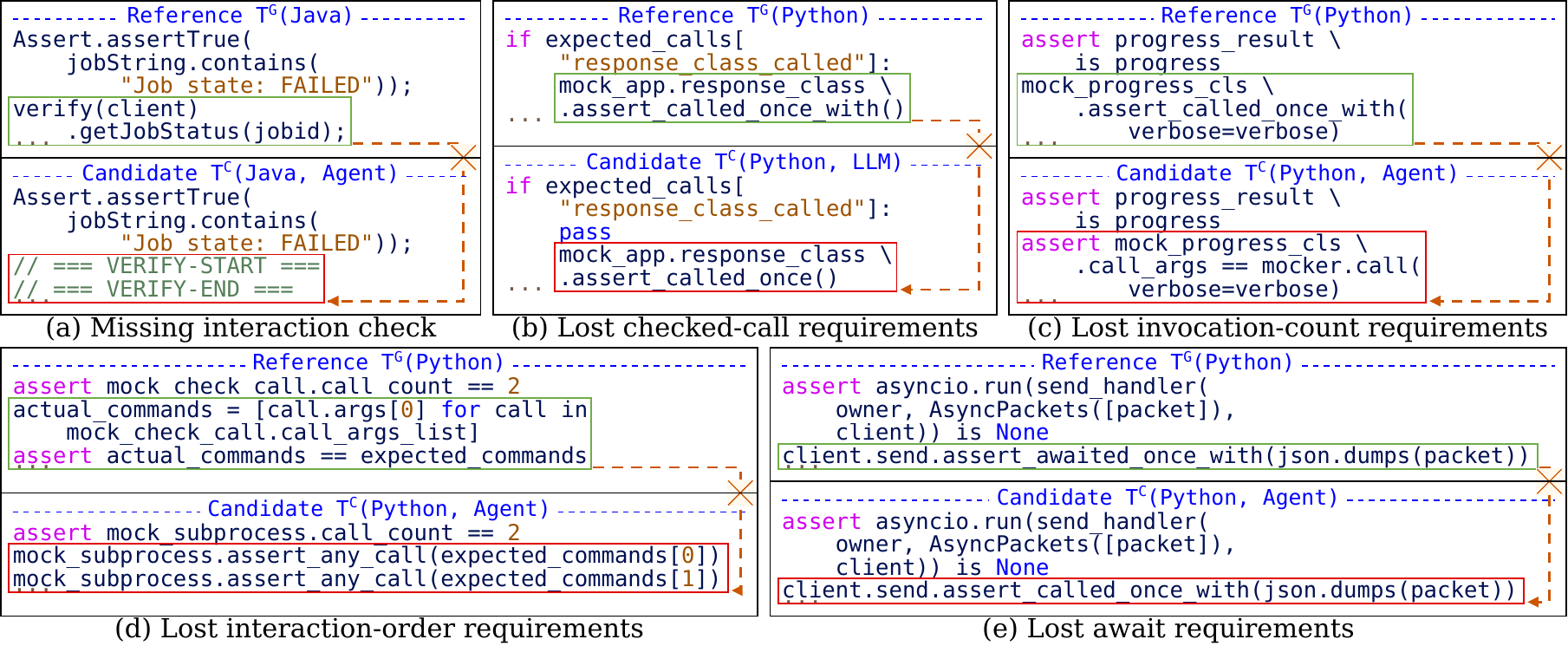}
\caption{Examples of unretained interaction check.}
\label{fig:reference-loss-assertions}
\Description{Five reference and candidate code comparisons showing
an omitted interaction check in Hadoop, a lost zero-argument
requirement in Flask, a lost exactly-once requirement in Langflow,
a lost command-order requirement in Crawl4AI, and a lost await
requirement in Tqsdk.
Reference requirements are framed in green, and corresponding
candidate fragments are framed in red.}
\end{figure}

\textbf{Mock assertion errors(C).}
\textbf{(1) Missing mock assertions(C-1):}
An interaction verified in $T^G$ is omitted entirely from $T^C$.
For example, in Hadoop's \texttt{testJobToString}, 
the agent completely drops the verification for intermediate dependency calls, including \texttt{getJobStatus}, \texttt{getTaskReports}, and \texttt{getTaskCompletionEvents}
(Figure~\ref{fig:reference-loss-assertions}(a)).
Consequently, although oracle assertions verify the final result, the critical dependency interactions enforced by $T^G$ remain unverified.
\textbf{(2) Call-matching errors(C-2):}
The interaction check is retained, but the constraints on its call arguments are weakened.
Specifically, in Flask(Figure~\ref{fig:reference-loss-assertions}(b)), the standalone candidate uses \texttt{assert\_called\_once()} for \texttt{response\_class}, preserving the reference call-count constraint while omitting the requirement that the call take no arguments.
\textbf{(3) Call-count verification errors(C-3):}
A check on call arguments is preserved, but the required invocation frequency is unenforced.
In Langflow, the candidate correctly inspects the arguments of the most recent invocation against the reference, but drops the assertion enforcing that it was invoked exactly once
(Figure~\ref{fig:reference-loss-assertions}(c)).
Validating the latest payload thus fails to guarantee that spurious intermediate calls did not occur.
\textbf{(4) Call-order verification errors(C-4):}
Individual invocations remain verified, but their execution sequence is disregarded.
While the Crawl4AI reference verifies the exact sequential execution of all actions against \texttt{expected\_commands}, the agent instead checks the two calls independently using unordered \texttt{assert\_any\_call} assertions
(Figure~\ref{fig:reference-loss-assertions}(d)).
Although this preserves call presence and multiplicity, it leaves the temporal order of side effects unchecked.
\textbf{(5) Missing await verification(C-5):}
An asynchronous interaction requirement is degraded into a synchronous invocation check.
In Tqsdk, the candidate substitutes the reference await assertion with a standard \texttt{assert\_called\_once\_with}
(Figure~\ref{fig:reference-loss-assertions}(e)).
Even though the payload and call count are verified, this substitution fails to ensure that the coroutine was actually scheduled and awaited before test completion.

\begin{findingbox}
\textbf{Finding 10:}
Standalone LLMs most frequently make \emph{stubbing errors} (68.75--71.43\%), whereas coding agents are dominated by \emph{mock assertion errors} (66.67--81.54\%), particularly missing assertions. In contrast, \emph{mock object errors} are consistently the least frequent (14.29--37.04\%).
\end{findingbox}

\subsection{Implications}\label{sec:implications}
\noindent\textbf{Implications for researchers:} (1) As the first systematic study on mock artifact generation, this work reveals a fundamental pitfall: LLMs and coding agents synthesize mocks merely to satisfy tests, so test execution status alone is a misleading quality signal and should not serve as the sole benchmark for assessment~\cite{barr2015oracle,zhu2025mockassertions}. (2) By jointly measuring executability, behavioral fidelity, and interaction-fault detection,  
we expose behavioral deviations and silent losses in fault detection that passing tests conceal, and offer ready-to-use evaluation criteria for future research. (3) Scaling repair rounds and time budgets substantially boosts executability, yet yields no commensurate gain in behavioral fidelity, evidencing that coding agents optimize for crash avoidance rather than semantic correctness. This implies that future techniques should embed interaction contracts directly into generation objectives~\cite{barr2015oracle,freeman2004mockroles,zhu2025mockassertions}. (4) Our failure analysis distills 660 failures into ten error types spanning mock objects, stubs, and assertions, all of which survive behind passing tests; this taxonomy converts previously invisible failure modes into explicit optimization targets for faithfully reconstructing mock behaviors and enforcing interaction checks.

\noindent\textbf{Implications for practitioners:} 
(1) We construct MockBench, the first benchmark for mock quality evaluation, comprising 165 validated mock-related samples from real-world Java and Python projects that can be directly adopted for industrial evaluation. (2) Our experiments across six LLMs and three coding agents reveal their heterogeneous strengths: agents excel at executability, but such gains do not translate into behavioral fidelity or fault detection, providing evidence-based guidance for technique selection in mock-based testing. (3) The sensitivity analysis shows that enlarging agent time budgets improves executability only within a limited range and never behavioral fidelity.
Practitioners should therefore treat passing tests as a necessary but insufficient checkpoint, invest in explicit behavioral verification rather than additional repair rounds, and consult our ten-type failure taxonomy to inspect the risks that remain concealed when generated tests pass~\cite{freeman2004mockroles,zhu2025mockassertions}.
\section{Threats to Validity}\label{sec:threats}
\textbf{Threats to external validity.} 
Our evaluation is limited to mock artifact generation in Java and Python, excluding other PLs. 
However, given that Java and Python represent both statically- and dynamically-typed PLs, carry mature mocking ecosystems, and are widely used in practical development, the threat is minimal.
Second, our findings may not generalize across all LLMs and agent architectures. These threats are also limited, as we tested diverse LLMs and coding agents of commercial, general, and code, as well as closed/open-source. Thus, experimental subjects are representative. 
As for the failure taxonomy, our extensive analysis spans two PLs, 660 distinct errors, four representative subjects (one LLM and three agents), and all three mocking artifacts. Hence, its generality threat is negligible.


\textbf{Threats to internal validity.} 
To prevent data leakage, MockBench features newly authored Python tests alongside Java tests augmented with additional mock artifacts. To reduce errors in manual construction, a lead reviewer examined all reference suites for dependency selection, scenario design, and mock-artifact completeness, followed by execution validation. Towards the subjectivity of failure analysis, each sample was independently inspected by two evaluators using a shared codebook developed through thematic analysis~\cite{cruzes2011recommended}, with disagreements resolved through discussion and expert adjudication. 
Hence, the above threats are minimal.


\section{Conclusion}\label{sec:conclusion}
This work presents MockBench, a bilingual benchmark for evaluating the generation of mock objects, stubs, and mock assertions. Through extensive experiments on six LLMs and three coding agents, we examine their capabilities in terms of executability, behavioral fidelity, and fault detection. The results show that coding agents significantly improve test executability, while passing tests may mask behavioral deviations and losses in interaction fault detection capability. Increasing the agents' time budget can improve executability within an effective range, but incurs higher resource costs without a corresponding improvement in behavioral fidelity. Finally, our systematic failure analysis identifies ten error types across three categories of mock artifacts, providing concrete directions for future research and practical guidance for reviewing generated mock artifacts.

\bibliographystyle{ACM-Reference-Format}
\bibliography{references}

@INPROCEEDINGS{7962389,
  author={Spadini, Davide and Aniche, Maurício and Bruntink, Magiel and Bacchelli, Alberto},
  booktitle={2017 IEEE/ACM 14th International Conference on Mining Software Repositories (MSR)},
  title={To Mock or Not to Mock? An Empirical Study on Mocking Practices},
  year={2017},
  volume={},
  number={},
  pages={402-412},
  doi={10.1109/MSR.2017.61},
  publisher = {IEEE}
}

@inproceedings{pan2025aster,
  title={Aster: Natural and multi-language unit test generation with llms},
  author={Pan, Rangeet and Kim, Myeongsoo and Krishna, Rahul and Pavuluri, Raju and Sinha, Saurabh},
  booktitle={2025 IEEE/ACM 47th International Conference on Software Engineering: Software Engineering in Practice (ICSE-SEIP)},
  pages={413--424},
  year={2025},
  organization={IEEE},
  doi = {10.1109/icse-seip66354.2025.00042},
  publisher = {IEEE}
}

@article{yang2026advancing,
  title={Advancing code coverage: Incorporating program analysis with large language models},
  author={Yang, Chen and Chen, Junjie and Lin, Bin and Wang, Ziqi and Zhou, Jianyi},
  journal={ACM Transactions on Software Engineering and Methodology},
  volume={35},
  number={5},
  pages={1--31},
  year={2026},
  publisher={ACM New York, NY}
}

@misc{hora2026coding,
  title={Are Coding Agents Generating Over-Mocked Tests? An Empirical Study},
  author={Hora, Andre and Robbes, Romain},
  year={2026},
  eprint = {2602.00409},
  archiveprefix = {arXiv},
  url = {https://arxiv.org/abs/2602.00409},
  howpublished = {arXiv preprint}
}

@inproceedings{qin2025mock,
  title={To Mock or Not to Mock: Divergence in Mocking Practices Between LLM and Developers},
  author={Qin, Hanbin},
  booktitle={2025 IEEE/ACM 47th International Conference on Software Engineering: Companion Proceedings (ICSE-Companion)},
  pages={239--240},
  year={2025},
  organization={IEEE},
  doi = {10.1109/icse-companion66252.2025.00077},
  publisher = {IEEE}
}

@inproceedings{arcuri2014automated,
  title={Automated unit test generation for classes with environment dependencies},
  author={Arcuri, Andrea and Fraser, Gordon and Galeotti, Juan Pablo},
  booktitle={Proceedings of the 29th ACM/IEEE international conference on Automated software engineering},
  pages={79--90},
  year={2014},
  doi = {10.1145/2642937.2642986},
  publisher = {ACM},
  address = {New York, NY, USA}
}

@inproceedings{saff2005automatic,
  title={Automatic test factoring for Java},
  author={Saff, David and Artzi, Shay and Perkins, Jeff H and Ernst, Michael D},
  booktitle={Proceedings of the 20th IEEE/ACM international Conference on Automated software engineering},
  pages={114--123},
  year={2005},
  doi = {10.1145/1101908.1101927},
  publisher = {ACM},
  address = {New York, NY, USA}
}

@inproceedings{saff2004mock,
  title={Mock object creation for test factoring},
  author={Saff, David and Ernst, Michael D},
  booktitle={Proceedings of the 5th ACM SIGPLAN-SIGSOFT workshop on Program analysis for software tools and engineering},
  pages={49--51},
  year={2004},
  doi = {10.1145/996821.996838},
  publisher = {ACM},
  address = {New York, NY, USA}
}

@inproceedings{10.1145/3510003.3510175,
author = {Fazzini, Mattia and Choi, Chase and Copia, Juan Manuel and Lee, Gabriel and Kakehi, Yoshiki and Gorla, Alessandra and Orso, Alessandro},
title = {Use of test doubles in Android testing: an in-depth investigation},
year = {2022},
isbn = {9781450392211},
publisher = {Association for Computing Machinery},
address = {New York, NY, USA},
url = {https://doi.org/10.1145/3510003.3510175},
doi = {10.1145/3510003.3510175},
booktitle = {Proceedings of the 44th International Conference on Software Engineering},
pages = {2266–2278},
numpages = {13},
location = {Pittsburgh, Pennsylvania},
series = {ICSE '22}
}

@INPROCEEDINGS{11025757,
  author={Li, Mengzhen and Fazzini, Mattia},
  booktitle={2025 IEEE/ACM 22nd International Conference on Mining Software Repositories (MSR)},
  title={DataTD: A Dataset of Java Projects Including Test Doubles},
  year={2025},
  volume={},
  number={},
  pages={851-855},
  doi={10.1109/MSR66628.2025.00124},
  publisher = {IEEE}
}

@inproceedings{mostafa2014empirical,
  title={An empirical study on the usage of mocking frameworks in software testing},
  author={Mostafa, Shaikh and Wang, Xiaoyin},
  booktitle={2014 14th international conference on quality software},
  pages={127--132},
  year={2014},
  organization={IEEE},
  doi = {10.1109/qsic.2014.19},
  publisher = {IEEE}
}

@misc{xu2026llm,
  title={LLM-based Low-Level Integration Test Generation for Java},
  author={Xu, Qinghua and Wang, Guancheng and Briand, Lionel and Guo, Zhaoqiang and Liu, Kui},
  year={2026},
  eprint = {2605.26851},
  archiveprefix = {arXiv},
  url = {https://arxiv.org/abs/2605.26851},
  howpublished = {arXiv preprint}
}

@misc{lee2026improving,
  title={Improving LLM-Driven Test Generation by Learning from Mocking Information},
  author={Lee, Jamie and Teh, Flynn and Zhu, Hengcheng and Li, Mengzhen and Fazzini, Mattia and Terragni, Valerio},
  year={2026},
  eprint = {2604.19315},
  archiveprefix = {arXiv},
  url = {https://arxiv.org/abs/2604.19315},
  howpublished = {arXiv preprint}
}

@inproceedings{zhu2020mocksniffer,
  title={Mocksniffer: Characterizing and recommending mocking decisions for unit tests},
  author={Zhu, Hengcheng and Wei, Lili and Wen, Ming and Liu, Yepang and Cheung, Shing-Chi and Sheng, Qin and Zhou, Cui},
  booktitle={Proceedings of the 35th IEEE/ACM international conference on automated software engineering},
  pages={436--447},
  year={2020},
  doi = {10.1145/3324884.3416539},
  publisher = {ACM},
  address = {New York, NY, USA}
}

@article{zhu2023stubcoder,
  title={{StubCoder}: Automated generation and repair of stub code for mock objects},
  author={Zhu, Hengcheng and Wei, Lili and Terragni, Valerio and Liu, Yepang and Cheung, Shing-Chi and Wu, Jiarong and Sheng, Qin and Zhang, Bing and Song, Lihong},
  journal={ACM Transactions on Software Engineering and Methodology},
  volume={33},
  number={1},
  pages={1--31},
  year={2023},
  publisher={ACM New York, NY, USA},
  doi={10.1145/3617171}
}

@article{nguyen2024automated,
  title={Automated test data generation and stubbing method for C/C++ embedded projects},
  author={Nguyen Tung, Lam and Binh Duong, Nguyen Vu and Le, Khoi Nguyen and Hung, Pham Ngoc},
  journal={Automated Software Engineering},
  volume={31},
  number={2},
  pages={52},
  year={2024},
  publisher={Springer}
}

@misc{bach2026library,
  title={Library-Aware Doubles and Iterative Repair for Large Language Model-Generated Unit Tests in OpenSIL Firmware},
  author={Bach, Ma Toan and Zheng, Yuchi and Razafindranto, Haingo and Alam, Tanvir and Leather, Aric and Sandhu, Ranveer and Arora, Jitesh},
  year={2026},
  eprint = {2606.19725},
  archiveprefix = {arXiv},
  url = {https://arxiv.org/abs/2606.19725},
  howpublished = {arXiv preprint}
}

@article{tiwari2024mimicking,
  title={Mimicking production behavior with generated mocks},
  author={Tiwari, Deepika and Monperrus, Martin and Baudry, Benoit},
  journal={IEEE Transactions on Software Engineering},
  volume={50},
  number={11},
  pages={2921--2946},
  year={2024},
  publisher={IEEE},
  doi={10.1109/TSE.2024.3458448}
}

@incollection{cheddadi2026google,
  title={Google test/google mock to verify critical embedded software},
  author={Cheddadi, Hafsa and Motahhir, Saad and Ghzizal, Abdelaziz},
  booktitle={Embedded Systems in Automotive Applications},
  pages={121--137},
  year={2026},
  publisher={Elsevier},
  doi = {10.1016/b978-0-443-33861-8.00009-4}
}

@inproceedings{nan2025test,
  title={Test intention guided llm-based unit test generation},
  author={Nan, Zifan and Guo, Zhaoqiang and Liu, Kui and Xia, Xin},
  booktitle={2025 IEEE/ACM 47th International Conference on Software Engineering (ICSE)},
  pages={1026--1038},
  year={2025},
  organization={IEEE},
  doi = {10.1109/icse55347.2025.00243},
  publisher = {IEEE}
}

@inproceedings{yang2024sweagent,
  title     = {{SWE}-agent: Agent-Computer Interfaces Enable Automated
               Software Engineering},
  author    = {John Yang and Carlos E. Jimenez and Alexander Wettig and
               Kilian Lieret and Shunyu Yao and Karthik R. Narasimhan
               and Ofir Press},
  booktitle = {Advances in Neural Information Processing Systems},
  year      = {2024},
  url       = {https://proceedings.neurips.cc/paper_files/paper/2024/file/5a7c947568c1b1328ccc5230172e1e7c-Paper-Conference.pdf},
  doi = {10.52202/079017-1601},
  pages = {50528--50652},
  publisher = {Curran Associates, Inc.},
  volume = {37}
}

@misc{anthropic2025claudecode,
  title={Claude Code},
  author={{Anthropic}},
  year={2025},
  howpublished={\url{https://github.com/anthropics/claude-code}},
  note={Accessed: 2026-06-01}
}

@misc{opencode2025,
  title={OpenCode: The Open Source {AI} Coding Agent},
  author={{OpenCode Contributors}},
  year={2025},
  howpublished={\url{https://github.com/opencode-ai/opencode}},
  note={Accessed: 2026-06-01}
}

@misc{openai2025gpt55,
  title={Using {GPT-5.5}},
  author={{OpenAI}},
  year={2026},
  howpublished={\url{https://developers.openai.com/api/docs/guides/latest-model?model=gpt-5.5}},
  note={Accessed: 2026-09-17}
}

@misc{anthropic2025sonnet46,
  title={Introducing {Claude Sonnet 4.6}},
  author={{Anthropic}},
  year={2026},
  howpublished={\url{https://www.anthropic.com/news/claude-sonnet-4-6}},
  note={Accessed: 2026-09-17}
}

@misc{qwen2025qwen37max,
  title={{qwen3.7-max} Model Info},
  author={{Alibaba Cloud}},
  year={2026},
  howpublished={\url{https://www.alibabacloud.com/help/en/model-studio/qwen3-7-max}},
  note={Accessed: 2026-09-17}
}

@misc{deepseek2025v4pro,
  title={{DeepSeek-V4-Pro} Model Card},
  author={{DeepSeek-AI}},
  year={2026},
  howpublished={\url{https://huggingface.co/deepseek-ai/DeepSeek-V4-Pro}},
  note={Accessed: 2026-09-17}
}

@misc{qwen2025qwen3coder,
  title={{Qwen3-Coder}: Official Model Repository},
  author={{Qwen Team}},
  year={2025},
  howpublished={\url{https://github.com/QwenLM/Qwen3-Coder}},
  note={Accessed: 2026-09-17}
}

@misc{junit2026userguide,
title={{JUnit} User Guide},
author={{JUnit Team}},
year={2026},
howpublished={\url{https://docs.junit.org/current/user-guide/}},
note={Accessed: 2026-06-27}
}

@misc{mockito2026,
title={{Mockito}},
author={{Mockito Team}},
year={2026},
howpublished={\url{https://site.mockito.org/}},
note={Accessed: 2026-06-27}
}

@misc{pytestmock2026,
title={{pytest-mock} Documentation},
author={{pytest-dev Team}},
year={2026},
howpublished={\url{https://pytest-mock.readthedocs.io/en/latest/}},
note={Accessed: 2026-06-27}
}

@misc{pytest2026,
title={{pytest} Documentation},
author={{pytest-dev Team}},
year={2026},
howpublished={\url{https://docs.pytest.org/en/stable/}},
note={Accessed: 2026-06-27}
}

@inproceedings{coles2016pit,
  author    = {Henry Coles and Thomas Laurent and Christopher Henard and Mike Papadakis and Anthony Ventresque},
  title     = {{PIT}: A Practical Mutation Testing Tool for {Java} (Demo)},
  booktitle = {Proceedings of the 25th International Symposium on Software Testing and Analysis},
  series    = {ISSTA 2016},
  pages     = {449--452},
  year      = {2016},
  publisher = {ACM},
  doi       = {10.1145/2931037.2948707},
  address = {New York, NY, USA}
}

@misc{cosmicray,
  author       = {{Sixty North AS}},
  title        = {{Cosmic Ray} Documentation},
  year         = {n.d.},
  howpublished = {\url{https://cosmic-ray.readthedocs.io/}},
  note         = {Mutation testing tool for Python. Accessed: 2026-09-28}
}

@article{zhang2025min,
  author  = {Peng Zhang and Zeyu Lu and Yang Wang and Yibiao Yang and Yuming Zhou and Mike Papadakis},
  title   = {Enriching Mutation Testing With Innovative Method Invocation Mutation: Filling the Crucial Missing Piece of the Puzzle},
  journal = {IEEE Transactions on Software Engineering},
  volume  = {51},
  number  = {7},
  pages   = {2125--2143},
  year    = {2025},
  doi     = {10.1109/TSE.2025.3573751}
}

@article{ahmed2025mocking,
  author    = {Mumtahina Ahmed and Md Nahidul Islam Opu and Chanchal Roy and Sujana Islam Suhi and Shaiful Chowdhury},
  title     = {Exploring Challenges in Test Mocking: Developer Questions and Insights from StackOverflow},
  journal   = {Journal of Systems and Software},
  year      = {2026},
  doi       = {10.1016/j.jss.2025.112748},
  note      = {Also available as arXiv:2505.08300},
  publisher = {Elsevier BV},
  pages = {112748},
  volume = {235}
}

@inproceedings{yu2019comprehending,
  author    = {Yu, Chak Shun and Treude, Christoph and Aniche, Mauricio Finavaro},
  title     = {Comprehending Test Code: An Empirical Study},
  booktitle = {2019 IEEE International Conference on Software Maintenance and Evolution (ICSME)},
  year      = {2019},
  pages     = {501--512},
  publisher = {IEEE},
  doi       = {10.1109/ICSME.2019.00084}
}

@inproceedings{cruzes2011recommended,
  author    = {Daniela S. Cruzes and Tore Dyb{\aa}},
  title     = {Recommended Steps for Thematic Synthesis in Software Engineering},
  booktitle = {2011 International Symposium on Empirical Software Engineering and Measurement},
  year      = {2011},
  pages     = {275--284},
  publisher = {IEEE},
  doi = {10.1109/esem.2011.36}
}

@inproceedings{wang-etal-2025-testeval,
    title = "{TestEval}: Benchmarking Large Language Models for Test Case Generation",
    author = "Wang, Wenhan  and
      Yang, Chenyuan  and
      Wang, Zhijie  and
      Huang, Yuheng  and
      Chu, Zhaoyang  and
      Song, Da  and
      Zhang, Lingming  and
      Chen, An Ran  and
      Ma, Lei",
    editor = "Chiruzzo, Luis  and
      Ritter, Alan  and
      Wang, Lu",
    booktitle = "Findings of the Association for Computational Linguistics: NAACL 2025",
    month = apr,
    year = "2025",
    address = "Albuquerque, New Mexico",
    publisher = "Association for Computational Linguistics",
    url = "https://aclanthology.org/2025.findings-naacl.197/",
    doi = "10.18653/v1/2025.findings-naacl.197",
    pages = "3547--3562",
    ISBN = "979-8-89176-195-7"
}

@inproceedings{widyasari2020bugsinpy,
  title = {BugsInPy: A Database of Existing Bugs in Python Programs to Enable Controlled Testing and Debugging Studies},
  author = {Widyasari, Ratnadira and Sim, Sheng Qin and Lok, Camellia and Qi, Haodi and Phan, Jack and Tay, Qijin and Tan, Constance and Wee, Fiona and Tan, Jodie Ethelda and Yieh, Yuheng and Goh, Brian and Thung, Ferdian and Kang, Hong Jin and Hoang, Thong and Lo, David and Ouh, Eng Lieh},
  booktitle = {Proceedings of the 28th ACM Joint Meeting on European Software Engineering Conference and Symposium on the Foundations of Software Engineering},
  year = {2020},
  doi = {10.1145/3368089.3417943},
  publisher = {ACM},
  address = {New York, NY, USA},
  pages = {1556--1560}
}

@misc{jain2024testgeneval,
  title = {TestGenEval: A Real World Unit Test Generation and Test Completion Benchmark},
  author = {Jain, Kush and Synnaeve, Gabriel and Rozière, Baptiste},
  year = {2024},
  eprint = {2410.00752},
  archivePrefix = {arXiv},
  primaryClass = {cs.SE},
  doi = {10.48550/arXiv.2410.00752},
  url = {https://arxiv.org/abs/2410.00752}
}

@article{spadini2019mock,
  title = {Mock objects for testing java systems: Why and how developers use them, and how they evolve},
  author = {Spadini, Davide and Aniche, Maurício and Bruntink, Magiel and Bacchelli, Alberto},
  year = {2019},
  journal = {Empirical Software Engineering},
  doi = {10.1007/s10664-018-9663-0},
  volume = {24},
  number = {3},
  pages = {1461--1498}
}

@inproceedings{arcuri2017mocking,
  title = {Private API Access and Functional Mocking in Automated Unit Test Generation},
  author = {Arcuri, Andrea and Fraser, Gordon and Just, Rene},
  year = {2017},
  booktitle = {2017 IEEE International Conference on Software Testing, Verification and Validation (ICST)},
  doi = {10.1109/icst.2017.19},
  pages = {126--137},
  publisher = {IEEE}
}

@article{zhu2025mockassertions,
  author  = {Hengcheng Zhu and Valerio Terragni and Lili Wei and Shing-Chi Cheung and Jiarong Wu and Yepang Liu},
  title   = {Understanding and Characterizing Mock Assertions in Unit Tests},
  journal = {Proceedings of the ACM on Software Engineering},
  volume  = {2},
  number  = {FSE},
  pages   = {554--575},
  year    = {2025},
  doi     = {10.1145/3715741}
}

@inproceedings{freeman2004mockroles,
  author    = {Steve Freeman and Tim Mackinnon and Nat Pryce and Joe Walnes},
  title     = {Mock Roles, Not Objects},
  booktitle = {Companion to the 19th Annual ACM SIGPLAN Conference on Object-Oriented Programming, Systems, Languages, and Applications},
  pages     = {236--246},
  year      = {2004},
  publisher = {ACM},
  doi       = {10.1145/1028664.1028765},
  address = {New York, NY, USA}
}

@article{jia2009higherorder,
  author  = {Yue Jia and Mark Harman},
  title   = {Higher Order Mutation Testing},
  journal = {Information and Software Technology},
  volume  = {51},
  number  = {10},
  pages   = {1379--1393},
  year    = {2009},
  doi     = {10.1016/j.infsof.2009.04.016}
}

@inproceedings{starace2025paperbench,
  title     = {{PaperBench}: Evaluating {AI}'s Ability to Replicate {AI} Research},
  author    = {Starace, Giulio and Jaffe, Oliver and Sherburn, Dane and Aung, James and Chan, Jun Shern and Maksin, Leon and Dias, Rachel and Mays, Evan and Kinsella, Benjamin and Thompson, Wyatt and Heidecke, Johannes and Glaese, Amelia and Patwardhan, Tejal},
  booktitle = {Proceedings of the 42nd International Conference on Machine Learning},
  pages     = {56843--56873},
  year      = {2025},
  volume    = {267},
  series    = {Proceedings of Machine Learning Research},
  publisher = {PMLR},
  url = {https://proceedings.mlr.press/v267/starace25a.html}
}

@inproceedings{chan2025mlebench,
  title     = {{MLE}-bench: Evaluating Machine Learning Agents on Machine Learning Engineering},
  author    = {Chan, Jun Shern and Chowdhury, Neil and Jaffe, Oliver and Aung, James and Sherburn, Dane and Mays, Evan and Starace, Giulio and Liu, Kevin and Maksin, Leon and Patwardhan, Tejal and Madry, Aleksander and Weng, Lilian},
  booktitle = {The Thirteenth International Conference on Learning Representations},
  year      = {2025},
  url = {https://arxiv.org/abs/2410.07095}
}

@article{ni2026gittaskbench,
  title   = {{GitTaskBench}: A Benchmark for Code Agents Solving Real-World Tasks Through Code Repository Leveraging},
  author  = {Ni, Ziyi and Wang, Huacan and Zhang, Shuo and Lu, Shuo and He, Ziyang and You, Wang and Tang, Zhenheng and Hu, Sen and Li, Bo and Hu, Chen and Jiao, Binxing and Jiang, Daxin and Du, Yuntao and Lyu, Pin},
  journal = {Proceedings of the AAAI Conference on Artificial Intelligence},
  volume  = {40},
  number  = {38},
  pages   = {32564--32572},
  year    = {2026},
  month   = {mar},
  doi     = {10.1609/aaai.v40i38.40533}
}

@article{barr2015oracle,
  author  = {Earl T. Barr and Mark Harman and Phil McMinn and Muzammil Shahbaz and Shin Yoo},
  title   = {The Oracle Problem in Software Testing: A Survey},
  journal = {IEEE Transactions on Software Engineering},
  volume  = {41},
  number  = {5},
  pages   = {507--525},
  year    = {2015},
  doi     = {10.1109/TSE.2014.2372785}
}

@article{jia2011mutation,
  author  = {Yue Jia and Mark Harman},
  title   = {An Analysis and Survey of the Development of Mutation Testing},
  journal = {IEEE Transactions on Software Engineering},
  volume  = {37},
  number  = {5},
  pages   = {649--678},
  year    = {2011},
  doi     = {10.1109/TSE.2010.62}
}

@inproceedings{just2014mutants,
  author    = {Ren{\'e} Just and Darioush Jalali and Laura Inozemtseva and Michael D. Ernst and Reid Holmes and Gordon Fraser},
  title     = {Are Mutants a Valid Substitute for Real Faults in Software Testing?},
  booktitle = {Proceedings of the 22nd ACM SIGSOFT International Symposium on Foundations of Software Engineering},
  pages     = {654--665},
  year      = {2014},
  publisher = {ACM},
  doi       = {10.1145/2635868.2635929},
  address = {New York, NY, USA}
}

@inproceedings{andrews2005mutation,
  author    = {Andrews, James H. and Briand, Lionel C. and Labiche, Yvan},
  title     = {Is Mutation an Appropriate Tool for Testing Experiments?},
  booktitle = {Proceedings of the 27th International Conference on Software Engineering},
  pages     = {402--411},
  year      = {2005},
  doi       = {10.1145/1062455.1062530},
  publisher = {ACM Press},
  address = {New York, New York, USA}
}

@incollection{papadakis2019mutation,
  author    = {Papadakis, Mike and Kintis, Marinos and Zhang, Jie and
               Jia, Yue and Le Traon, Yves and Harman, Mark},
  title     = {Mutation Testing Advances: An Analysis and Survey},
  booktitle = {Advances in Computers},
  volume    = {112},
  pages     = {275--378},
  publisher = {Elsevier},
  year      = {2019},
  doi       = {10.1016/bs.adcom.2018.03.015}
}

@misc{apacheMaven,
  year         = {2026},
  author       = {{Apache Software Foundation}},
  title        = {Apache Maven},
  howpublished = {\url{https://maven.apache.org/}},
  note         = {Accessed: 2026-10-02}
}

@misc{gradle,
  year         = {{[n.d.]}},
  author       = {{Gradle Inc.}},
  title        = {Gradle User Manual},
  howpublished = {\url{https://docs.gradle.org/current/userguide/}},
  note         = {Accessed: 2026-10-02}
}

\end{document}